# The Natural Science Underlying Big History

Eric J. Chaisson

**Harvard-Smithsonian Center for Astrophysics**

**Harvard University, Cambridge, Massachusetts 02138 USA**

ejchaisson@cfa.harvard.edu

**Abstract**

Nature's many varied complex systems—including galaxies, stars, planets, life, and society—are islands of order within the increasingly disordered Universe. All organized systems are subject to physical, biological or cultural evolution, which together comprise the grander interdisciplinary subject of cosmic evolution. A wealth of observational data supports the hypothesis that increasingly complex systems evolve unceasingly, uncaringly, and unpredictably from big bang to humankind. This is global history greatly extended, big history with a scientific basis, and natural history broadly portrayed across ~14 billion years of time. Human beings and our cultural inventions are not special, unique, or apart from Nature; rather, we are an integral part of a universal evolutionary process connecting all such complex systems throughout space and time. Such evolution writ large has significant potential to unify the natural sciences into a holistic understanding of who we are and whence we came.

No new science (beyond frontier, non-equilibrium thermodynamics) is needed to describe cosmic evolution's major milestones at a deep and empirical level. Quantitative models and experimental tests imply that a remarkable simplicity underlies the emergence and growth of complexity for a wide spectrum of known and diverse systems. Energy is a principal facilitator of the rising complexity of ordered systems within the expanding Universe; energy flows are as central to life and society as they are to stars and galaxies. In particular, energy rate density—contrasting with information content or entropy production—is an objective metric suitable to gauge relative degrees of complexity among a hierarchy of widely assorted systems observed throughout the material Universe. Operationally, those systems capable of utilizing optimum amounts of energy tend to survive, and those that cannot are non-randomly eliminated.

## 1. Introduction

For many years, my scientific research has explored natural science broadly yet deeply, striving to place humanity into a cosmological framework. I have especially sought to analyze the apparent rise of complexity among principal, organized systems throughout the ~14-billion-year-old Universe—mainly galaxies, stars, planets, life, and society—as well as to decipher the myriad evolutionary events that have produced intelligent beings and their technological machines on Earth. In doing so, I have ventured beyond mere words and subjective analyses, rather to strongly and objectively embrace empirical findings while using hard-science methodology to quantitatively synthesize understanding.

Some scholars [1-5] call this subject "big history," wherein they trace a chronicle of events and systems that helped produce specifically us: the Milky Way Galaxy, the Sun, the Earth, and human beings. The result is a compelling, yet provincial, narrative of our past and present—an attempt to relate specifically how humanity emerged within a long and remarkable story spanning unusually deep time. Somewhat by contrast, when considering the Universe at large, including all such galaxies and stars, and not least the possibility of other Earth-like planets replete with potential intelligent beings, I have always referred to this subject more inclusively as "cosmic evolution" [6-13]. This is interdisciplinary natural history writ large—a broad synthesis of natural science with an equally broad definition, to wit: *Cosmic evolution is the study of the many varied developmental and generational changes in the assembly and composition of radiation, matter, and life throughout the history of the Universe.*

I have recently reviewed aspects of cosmic evolution in technical journals [14-18] and produced several books and films on the subject [9, 19], as well as taught the subject in university classrooms for decades [20, 21]. All these materials are widely available in the published literature and easily accessible on-line, as noted in the references at the end of this paper, which is itself both a comprehensive account of the subject to date and a prequel to a subsequent paper that examines several practical applications of cosmic evolution to global issues now confronting humankind on Earth [22], including ways that this interdisciplinary synthesis might both explain and guide our human condition now and in the future.

This article provides not only an underlying scientific rationale but also data-rich insights for the newly emerging subject of big history. In contrast to the established discipline of world (or global) history, which is limited to the study of relatively modern humanity on our particular planet, hence relates mainly to the recent past recorded in writing, big history extends much farther back in time. Similar to yet even broader than what our forebears called natural history, big history seeks to understand (indeed to emphasize) humankind within the larger context of truly deep time. It explores our remote roots that extend literally into the wider cosmos, from elementary particles of the early

Universe to cultured life on planet Earth, but it restricts its purview mostly to phenomena pertinent to specifically our Milky Way, our Sun, our Earth, and ourselves. Big historians, like all historians, basically strive to know themselves—nobly and ideally, yet sometimes dubiously rendering humanity as central or special while deciphering our sense of place in the grand scheme of things ("…human history in its wider context" [1], "…human history within the context of cosmic history" [3], or "…human history as part of a much larger story" [5], alas an historical approach often allied with a poetic expression that "…the proper study of mankind is man" [23].

In my own research, I further distinguish big history from cosmic evolution, which also has aliases of cosmological history, universal history, epic of evolution, and sometimes astrobiology; the former chronicles events mainly relevant to the advent and exploits of humanity whereas the latter adopts a more general purview regarding the origin, evolution, and fate of all galaxies, stars, planets, and life throughout the expansive and expanding Universe. My interests focus neither solely on human and planetary history nor even merely on cosmic history regarding humanity; rather, I aim to explicate a broad cosmic narrative that includes our own big history as part of an overarching universal worldview. Thus, as I have argued elsewhere [24, 25] and continue to do so here, cosmic evolution is a more ambitious undertaking than big history; cosmic evolution relates specific evolutionary actions within a more general synthesis of myriad changes that likely produced all material things. To be sure, cosmic evolutionists regard humankind as a miniscule segment of an extraordinarily lengthy story, in fact a tiny strand that enters only in the most recent ~0.01% of the story to date—akin to an uber-movie of 14 billion years that plays linearly for 14 minutes, yet in which humankind appears well within the last second of the film [19]. Even so, it is the scientifically oriented cosmic-evolutionary scenario described below that technically bolsters the humanistically oriented big-history enterprise with rigorous, quantitative natural science. In turn, we can learn a great deal about cosmic evolution in general by studying the principal complexifying stages and its underlying processes that created us in particular. What follows here therefore, in this empirical analysis of big history per se, is a limited examination of some of the many salient evolutionary events that gave rise to increasingly complex systems along an aimless, meandering path leading eventually and remarkably to humankind on Earth.

## 2. A Grand Evolutionary Synthesis

The past few decades of scientific research have seen the emergence of a coherent description of natural history, including ourselves as intelligent beings, based on the ancient concept of change. Heraclitus may have been right 25 centuries ago when he made perhaps the best observation of Nature ever: $\pi\alpha\nu\tau\alpha\ \rho\epsilon\iota$, translated variously as "all flows, all fluxes, or nothing stays [the same]." From stars and galaxies to life and humanity, a growing scholarly community is now discovering an intricate pattern of understanding throughout all the sciences—an interdisciplinary story of the origin and evolution of every known type of object in our richly endowed Universe. The result is a grand evolutionary synthesis linking a wide variety of academic specialties—physics, astronomy, geology, chemistry, biology, anthropology, among others and including social studies and the humanities as well—a cosmological epic of vast proportions extending from the very beginning of time to the present—and presumably on into the future.

Given the new intellectual age of interdisciplinarity, we are beginning to decipher how all known systems—atoms and galaxies, cells and brains, people and society, among innumerable others—are interrelated and constantly changing. Our appreciation for evolution now extends well beyond the subject of biology; the concept of evolution, generally considered (as in most dictionaries) as *any process of ascent with change in the formation, growth, and development of systems*, has become a robust unifying factor within and among all of the sciences. Yet questions remain: How realistic is our search for unity in Nature, and will the integrated result resemble science or philosophy? How have the magnificent examples of order on and beyond Earth arisen from chaos? Can the observed constructiveness of cosmic evolution be reconciled with the inherent destructiveness of thermodynamics? Most notably, what processes underlie the origin and evolution of so many diverse structures spanning the Universe and especially their growing complexity as defined by *intricacy, complication, variety, or involvement among the interconnected parts of a system*?

Recent research, guided by huge new databases detailing a multitude of complex systems, offers rational answers to some of the above questions. Growing order within "islands" of complexity such as galaxies, stars, planets, life, and society is outpaced by great "seas" of increasing disorder elsewhere in the environments beyond those systems. All such complex systems quantitatively obey the valued precepts of modern thermodynamics, especially frontier non-equilibrium thermodynamics. None of Nature's organized structures, not even life itself, is a violation (nor even a circumvention) of the celebrated $2^{nd}$ law of thermodynamics. Both order and entropy can increase together—the former locally (in systems) and the latter globally (in surrounding environments). Thus, we arrive at a central question lurking in the minds of some of today's eclectic thinkers (*e.g.*, [26-29]): Might there be a kind of essential Platonism at work in the Universe—a general principle, a unifying law, or perhaps a surprisingly simple process that naturally creates, organizes, and maintains the form and function of complex systems everywhere?

Figure 1 depicts an archetypal illustration of cosmic evolution—the arrow of time—extending from big bang to humankind. Regardless of its shape or orientation, such an arrow represents a symbolic guide to the sequence of events that have changed systems throughout all of history from simplicity to complexity, from inorganic to organic, from chaos to order. That sequence, as determined by a large amount of data collected since Renaissance times, accords well with the idea that a thread of change links the evolution of primal energy into elementary particles, after which those particles changed into atoms, in turn those atoms collected into galaxies and stars that then fused the heavy elements, followed by the evolution of those elements into the molecular building blocks of life, of those molecules into life itself, and of intelligent life into the cultured and technological society that we humans now comprise. Despite the specialization of today's academic research, evolution crosses all disciplinary boundaries. As such, the most familiar kind of evolution—biological evolution, or neo-Darwinism—is just one, albeit important, subset of broader evolutionary action encompassing much more than mere life on Earth. In short, what Darwinian change does for plants and animals, cosmic evolution aspires to do for all material systems. And if Darwinism created a revolution of understanding that humans are no different from other life-forms on our planet, then cosmic evolution extends the simple, yet powerful, idea of change writ large by treating Earth and our bodies in much the same way as stars and galaxies far beyond.

*Cosmic Evolution:*

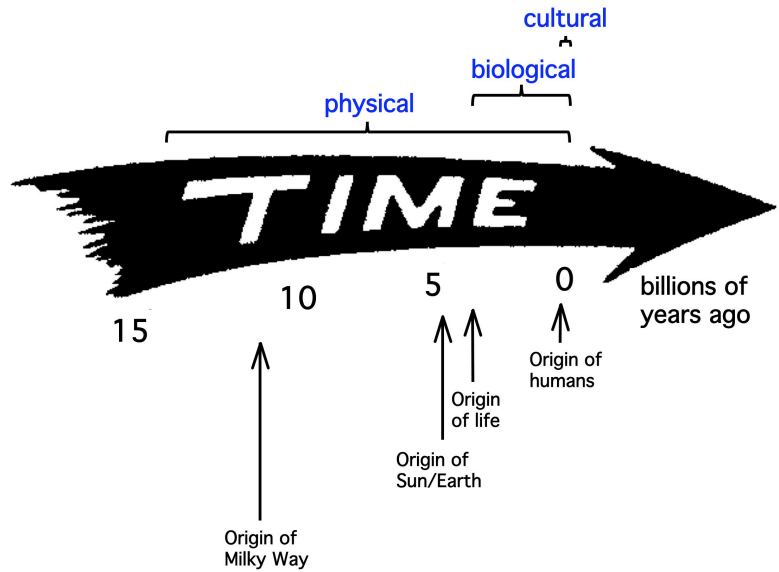

FIGURE 1: Extending over ~14 billion years from the big bang (left) to the present (right), an "arrow of time" symbolically represents the grand sweep of cosmic evolution, an interdisciplinary synthesis of all the natural sciences. Cosmic evolution generally integrates the three phases of physical, biological, and cultural evolution (top), and specifically includes the "big history" of our galaxy, star, and planet, as well as of life, humanity, and civilization (bottom). Despite this arrowhead sketch, there is no directionality implied for the evolutionary process, nor any purpose, plan, or design evident in the data supporting it.

Anthropocentrism is neither intended nor implied by the arrow of time; there is nothing directional about it. Aimed only toward the future, this graphical arrow points at nothing particular in space, and certainly not humanity. Anthropic principles notwithstanding, no logic or data support the idea that the Universe was conceived to produce specifically us [30-33]. And although humans and our cultural achievements dominate discourse among big historians (*e.g.*, [5]), no evidence implies that we are the pinnacle or culmination of the cosmic-evolutionary scenario (even though some biological systems per se may be nearing their complexity limits [34])—nor are we likely the only technologically competent beings who have emerged in the organically rich Universe.

Time's arrow merely provides a convenient symbol, artistically depicting ubiquitous changes that have produced increasingly complex structures from spiral galaxies to rocky planets to thinking beings. Nor does the arrow express or imply that "lower," primitive life-forms biologically change directly into "higher," advanced organisms, any more than galaxies physically change into stars, or stars into planets. Rather, with time—much time—the environmental conditions suitable for spawning simple life eventually changed into those favoring the emergence of more complex species; likewise, in the earlier Universe, environments ripened for galactic formation, but now those conditions are more conducive to stellar and planetary formation. Changes in surrounding environments, especially their energy budgets, often precede the evolution of ordered systems, and the resulting system changes have *generally* been toward greater amounts of diverse complexity, as numerically condensed in the next section.

## 3. Complexification via Energy Flows

My research agenda in cosmic evolution attempts to interpret natural history over many billions of years, and to do so by embracing a fundamental leitmotif of energy flow through increasingly complex systems. This is not a criticism of colleagues who examine complexity and evolution by employing information theory or entropy production, although I personally find these methods overly abstract (with dubious intentions), hard to define (to everyone's satisfaction), and even harder to measure (on any scale). Regarding the latter, neither maximum nor minimum entropy principles are evident in the data presented in this review. Regarding the former, I sense, but cannot prove, that information is another kind of energy; both information storage and retrieval require energy, and greater information processing and calculation need high energy density. While information content and entropy value are useful terms that offer theoretical insight, neither provides clear, unambiguous, empirical metrics. At least one leading researcher recently advised they be "banned from interdisciplinary discussions of complexity in the history of the Universe" [35]. As an experimental physicist, I sense that information may aid the description of some systems, but that energy is needed in the creation or operation of all of them.

Notwithstanding their taxing, controversial semantics, entropy production [36, 37] and information content [38-41] are frequently espoused in discussions of origin, evolution, and complexity. Yet, these alternative methods of diagnosing systems are less encompassing and less empirical than many researchers admit, their theoretical usefulness narrow, qualitative, and equivocal in deciphering, or even characterizing, a topic as promising as authentic complexity science. Although yielding fruitful properties of systems and their emergent and adaptive qualities unlikely to be understood otherwise, such efforts have reaped an unusual amount of controversy and only limited success to date [42]. Nor are information or negentropy practically useful in quantifying or measuring complexity. In biology alone, much as their inability to reach consensus on life's definition, biologists cannot agree on a complexity metric. Some use numerical genome size [43], others gauge body morphology and functional flexibility [44], still others count cell types in organisms [27, 45], chart cellular specialization among species [46], or appeal to networks of ecological interactions [47]. Some of these attributes of life have qualitative worth, yet few hold quantitatively. For example, among morphologically primitive organisms, such as sponges and pre-metazoans, meager cell types often differ dramatically with their genomic wealth [34]. Furthermore, humans' 3.2 billion base pairs well exceed that of a pufferfish (~365 million) yet are greatly exceeded by closely related lungfish (~133 billion), and even the wheat genome, which is arguably the most important plant to humans, is at ~17 gigabases several times the size of our human genome; likewise humans' ~22,300 genes are dwarfed by the ~33,000 genes in a scorpion, ~37,000 in a banana, and ~57,000 in an apple. It is time to retreat from information-based and type-counting complexity metrics; protein-coding genes and their base pairs might serve to characterize genomes, but they are faulty markers of species complexity.

The Universe is not likely an information-wired machine obeying a fixed computer program. Rather, the vast and changeful cosmos seems to be an arena for evolution, as a winding, rambling process that includes both chance and necessity, to produce a wide spectrum of ordered, organized systems over the course of very long periods of historical time. Such frequent, ongoing, ubiquitous change seems nothing more (yet nothing less) than the natural way that cultural evolution developed beyond biological evolution, which in turn built upon physical evolution before that. Each of these evolutionary phases comprises an integral part of cosmic evolution's larger purview that also operates naturally, as it always has and likely always will, with the irreversible march of time in the expanding Universe.

Cosmic evolution as understood today is governed largely by the laws of physics, especially those of thermodynamics. Note the adverb "largely," for this is not an exercise in traditional reductionism. Of all the known principles of Nature, thermodynamics perhaps best describes the process of

change—yet change dictated by a combination of both randomness and determinism. Literally, thermodynamics, which specifies what can happen not what necessarily will happen, connotes "movement of heat"; a more insightful translation (in keeping with dynamics implying change generally) would be "change of energy." Energy flows caused by the expanding cosmos do seem to be as central and common to the structure and function of all complex systems as anything yet discovered in Nature. Furthermore, the optimized use of such energy flows by complex systems, as argued below, might well act as a motor of cosmic evolution on larger scales, thereby affecting physical, biological, and cultural evolution on smaller scales.

The idea that energy is at the heart of all material things is not new. Again it was Heraclitus, noted above as the ancient world's foremost champion of widespread change in Nature, who may have best appreciated the cause of all that change. The etymology of the term "energy" dates back to ~500 BCE, when this "philosopher of flux and fire" used the word *en-ergon* to describe "the father of everything . . . and the source of all activity" [48]. Credit is fair where credit is due, even if this Greek thinker was apparently disinclined to test his ideas with empirical, quantitative analyses that are fundamental to our modern scientific methods.

Energy not only plays a role in ordering and maintaining complex systems; it might also determine their origin, evolution, and destiny. Recognized decades ago at least qualitatively in words and mostly in biology [49-51], the need for energy is now embraced as an essential organizing feature not only of biological systems such as plants and animals but also of physical systems such as stars and galaxies (*e.g.*, [52-58]). If fusing stars had no energy flowing within them, they would collapse; if plants did not photosynthesize sunlight, they would shrivel up and die; if humans stopped eating, we too would perish. Energy's central role is also widely recognized in cultural systems such as a city's inward flow of food and resources amidst its outward flow of products and wastes; indeed, energy is key to today's economy, technology, and civilization [22]. All complex systems—whether alive or not—are open, organized, non-equilibrated structures that acquire, store, and utilize energy.

Energy, therefore, is a quantity that has commonality among many complex systems and not least considerable appeal to physical intuition—a classic term that is well definable, understandable, and above all measurable. Even so, the quantity of choice cannot be energy alone, for a star is clearly more energetic than a flower, a galaxy much more energetic than a single cell. Yet any living system is surely more complicated than any inanimate entity. Absolute energies are not as indicative of complexity as relative values, which depend on a system's size, composition, coherence, and function. To characterize complexity objectively—that is, to normalize all such structured systems in precisely the same way—a kind of energy density is judged most useful. Moreover, it is the *rate* at which (free) energy transits complex systems of given mass that seems especially constructive (as has long been realized for ecosystems: [49, 59-60], thereby delineating energy *flow*. Hence, "energy rate density" (also termed power density), symbolized by $\Phi_m$, is a useful operational term whose expressed intent and plain units are easily understood, indeed whose definition is clear: *the amount of energy passing through a system per unit time and per unit mass*. In this way, neither new science nor mystical appeals to non-science are needed to explain the impressive hierarchy of complex systems in the cosmic-evolutionary narrative, from quarks to quasars, from microbes to minds.

Cosmic evolutionists are now expanding and deepening our knowledge of evolution in the broadest sense; we seek to push the analytical envelope beyond mere words, in fact beyond biology. Specifically, as explained in this review, we use aspects of energy to quantitatively decipher much of big history. Experimental data and detailed computations of energy rate densities are reported elsewhere [16, 17], most of them culled or calculated from values published in widely scattered journals over many years. Here is the briefest of compact summaries, whose ranked contents will be further examined and critiqued in subsequent sections of this review:

- Among physical systems, stars and galaxies generally have energy rate densities ($10^{-3}$ - $10^2$ erg/s/g) that are among the lowest of known organized structures. Galaxies display temporal trends in rising values of $\Phi_m$ while developing, such as for our Milky Way, which increased from ~$10^{-2}$ to 0.1 erg/s/g while changing from a primitive dwarf galaxy into a mature spiral galaxy. Stars, too, adjust their internal states while evolving during one or more generations, their $\Phi_m$ values rising while complexifying with time as their interior thermal and chemical gradients steepen and differentiate; for the Sun, $\Phi_m$ increases from ~1 to $10^2$ erg/s/g while changing from a young protostar to an aged red giant.

- In turn, among biological systems, plants and animals regularly exhibit intermediate values of $\Phi_m = 10^3$ - $10^5$ erg/s/g. For plant life on Earth, energy rate densities are well higher than those of normal stars and typical galaxies, as perhaps best demonstrated by the evolution of photosynthesizing gymnosperms, angiosperms, and $C_4$ plants, which over the course of a few hundred million years increased their $\Phi_m$ values nearly an order of magnitude to ~$10^4$ erg/s/g. Likewise, as animals evolved from fish and amphibians to reptiles, mammals, and birds, their $\Phi_m$ values rose still more, from ~$10^{3.5}$ to $10^5$ erg/s/g. Energy conceivably acted as a mechanism of change, partly and optimally selecting systems able to utilize increased power densities, while forcing others to destruction and extinction—all likely in accord with the widely accepted Darwinian principles of biological selection. Not surprisingly, brains have among the highest values of $\Phi_m$ for all living things.

- Furthermore, for cultural systems, advances in technology are comparable to those of human society itself, each of them energy-rich and having $\Phi_m \geq 10^5$ erg/s/g—hence plausibly among the most complex systems known. Social evolution can be tracked, again in terms of normalized energy consumption, for a variety of human-related cultural advances among our ancestral forebears, from early agriculturists (~$10^5$ erg/s/g) to modern technologists (~$10^{6.5}$). Machines, too, and not just computers, but also ordinary engines that drive today's economy, show the same upward trend from primitive devices of the industrial revolution (~$10^5$ erg/s/g) to today's sophisticated jet aircraft (~$10^{7.5}$).

Of special note, although the absolute energy in astronomical systems greatly exceeds that of our human selves, and although the mass densities of stars, planets, bodies, and brains are all comparable, the energy rate densities for human beings and our modern society are approximately a million times greater than for stars and galaxies. That's because the quantity $\Phi_m$ is an energy rate *density*. For example, although the Sun emits much luminosity, $4 \times 10^{33}$ erg/s (equivalent to nearly a billion billion billion Watt lightbulb), it also contains an unworldly large mass, $2 \times 10^{33}$ g; thus each second an amount of energy equaling only 2 ergs passes through each gram of this star. In contrast to any star, more energy (thousands of ergs) flows through each gram of a plant's leaf during photosynthesis, and much more energy (nearly a million ergs) pervades each gram of gray matter in our brains while thinking.

Figure 2, which is plotted on the same temporal scale as in Figure 1, graphically compiles those data compactly presented in the three bullets above, thereby depicting in a single plot the increase of $\Phi_m$ as measured or computed for representative systems that emerged at widely different times in natural history. (For specific power units of W/kg, divide $\Phi_m$ by $10^4$.) This "master graph" not only encapsulates on one page the physical, biological, and cultural evolution of homogeneous, primordial matter of the early Universe into organized systems of increased intricacy and energy rate density, but also shows how evolution has done so with increasing speed, hence the exponentially rising curve. The $\Phi_m$ values and historical dates plotted here are estimates for the general category to which each system belongs, yet variations and outliers are inevitable, much as expected for any simple, unifying précis of a messy, imperfect Universe. It is not the precise values of these many plotted quantities that matter most as much as the generally upward trend of $\Phi_m$ with the passage of time.

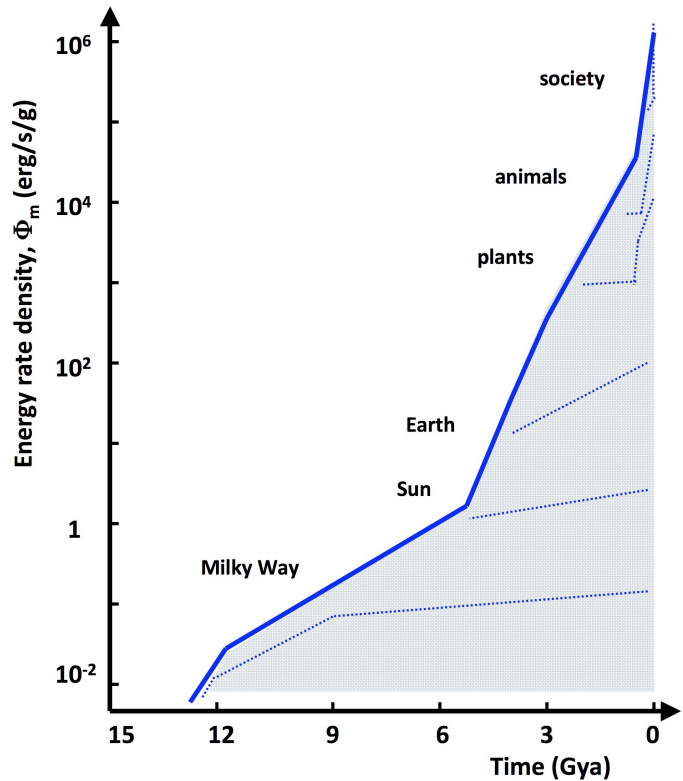

FIGURE 2: Energy rate density, $\Phi_m$, for a wide spectrum of systems observed throughout Nature displays a clear increase across ~14 billion years, implying rising complexity throughout all known historical time. The solid blue curve in this "master plot," graphed on the same temporal scale as in Figure 1, implies an exponential rise as cultural evolution (steepest slope at upper right) acts faster than biological evolution (moderate slope in middle part of curve), which in turn surpasses physical evolution (smallest slope at lower left). The shaded area includes a huge ensemble of $\Phi_m$ values as many different individual types of complex systems continued changing and complexifying since their origin; the several small dotted blue lines within that shaded area delineate some major evolutionary events that are then graphed in greater detail in Figures 3-9. The $\Phi_m$ values and historical dates plotted here are estimates for specific systems on the evolutionary path that led to humankind—namely, the Galaxy, Sun, and Earth, as well as much life all across our planet. As such, this particular graph is of greatest relevance to big historians seeking to understand how human society emerged naturally over the course of all time.

Energy is apparently a common currency for all complex, ordered systems. Even for structures often claimed to be "self-assembled" or "self-organized," energy is inexorably involved, as noted in the clarifying discussion in Sec. 5.1. Energy flow is among the most unifying processes in all of science, helping to provide cogent explanations for the origin, evolution, and complexification of a vast array of systems spanning >20 orders of magnitude in scale and nearly as many in time—notably, how systems emerge, mature, and terminate during a single generation as well as across multiple generations. Big historians have quickly embraced the centrality of energy in evolutionary events that yielded greater complexity, even if their interpretations and classifications sometimes differ from one another [61-63, 5].

Robust systems, whether stars, life, or civilization, have optimum ranges of energy flow; too little or too much and systems abort. Optimality is likely favored in the use of energy—a concept that I have long emphasized (*e.g.*, [8]) and as stressed further in Sec. 5.3—not so little as to starve a system, yet not so much as to destroy it. The data communicated below show no maximum energy principles, minimum entropy states, or entropy production criteria [49, 64-66]. Better metrics might describe each of the individual systems governed by physical, biological, and cultural evolution, but no other metric seems capable of uniformly describing them altogether. The significance of plotting "on the same page" (as literally done in Figure 2) a single empirical quantity for such an extraordinarily wide range of complex systems observed throughout Nature should not be underestimated.

## 4. The Principal Systems of Big History

*4.1. Milky Way Galaxy*

*4.1.1. Origin and Evolution of the Milky Way.* Although we cannot look directly into the past and watch our own Galaxy forming and evolving, we can study other, similar systems, including their basic building blocks. The following account of a widely accepted scenario for the origin and evolution of the Milky Way Galaxy, minus lingering, controversial details, explains much of its galactic structure observed today as well as the kinematical and chemical properties of its stellar populations [67-71]. In the main, our Galaxy (conventionally written with a capital "G" to distinguish our own such system, the Milky Way, from myriad others) resembles a "cannibal" that consumed at least hundreds of smaller galaxies or galactic fragments during its "lifetime" to date. The great majority of the Galaxy likely originated within the Universe's first 1-4 billion years (Gy) by means of dynamic, non-equilibrium mergers among several smaller systems, each of them contracting pregalactic clumps of mostly dark matter having masses $\sim 10^{7-8}$ $M_\odot$ (where the Sun's mass, 1 $M_\odot \approx 2 \times 10^{33}$ g)—comparable to the smallest dwarf galaxies and the biggest globular clusters, all of which have low heavy-element abundance implying ancient formation from relatively unprocessed gas. Today's few-dozen dwarf galaxies in the Local Group (our parent galaxy cluster) are probably surviving remnants of those immature massive fragments that have not yet merged with the Milky Way [72]; and the ~160 known globular clusters in its halo may be archaic fossils (gravitationally stripped cores) of some of those dwarfs galaxies that did merge [73].

Initially an irregular region $\sim 10^5$ light-years in diameter whose oldest stars now (mostly in the halo)

outline that birth, the Galaxy's baryonic gas and dust eventually settled into a thin spinning disk whose dimensions roughly match those measured today and where abundant young stars are found among others still forming.  Timescales for subsequent evolution during the past ~10 Gy wherein the Galaxy's size, shape, and composition were altered are still debated, although a recently discovered thick (~6x10$^3$ light-year) disk containing middle-aged stars (7-10 Gy old; ~0.5% elements heavier than He) may represent an intermediate stage of star formation that occurred while the gas was still falling into the thinner plane.  It also remains unclear if the original galactic building blocks contained already formed, even older (0% heavy-element) stars or if they resembled (and may still include) the dwarf galaxies seen today, some of which do have stars, others merely atomic gas.  In any case, such hierarchical clustering of dark matter clumps provides the conceptual framework for modern studies of galaxy evolution, describing a process of upward assembly that began many billion years ago and continues, albeit at greatly reduced rate, to the present [74, 75].

Studies of the composition of stars in the galactic disk suggest that the infall of halo gas is still occurring today; the star-forming lifetime of a spiral disk may be prolonged by the arrival of fresh gas from the Galaxy's surroundings.  However, it is unlikely that any major mergers ever impacted our Milky Way, otherwise its fragile thin disk would not have survived.  Models of star formation and stellar nucleosynthesis imply that the fraction of heavy elements in disk stars should be significantly greater than observed, unless the gas in the disk is steadily diluted by relatively pristine gas arriving from the halo (or beyond) at rates of 5-10 M$_\odot$/y.  Recently discovered in the galactic halo are several streams of stars with similar orbits and compositions, each thought to be remnants of dwarf galaxies torn apart by the Galaxy's tidal field and eventually "digested" by our Galaxy, much as other dwarf companion galaxies were probably "consumed" by it long ago [76].  The small Sagittarius dwarf galaxy (~10$^9$ M$_\odot$), the closest member of the Local Group now approaching the center of the Milky Way's far side, has been experiencing its death throes for the past ~3 Gy and will likely be assimilated into the Milky Way within another 1 Gy [77]; simulations imply that the Magellanic Clouds will eventually meet the same fate [78].  Upwards of a thousand mini-galaxies must have been likewise captured, shredded, and dissolved into the formative Milky Way long ago, their stellar inhabitants now intermingling with our Galaxy's indigenous population.  Such galactic archaeology is supported by recent observations of the nearby Andromeda galaxy, where relics of past cannibalism between it and its satellite dwarf galaxies (notably filamentary streams of stars in its halo) show the hierarchical process at work [79].

Astronomers have long suspected that galaxies sustain themselves by acquiring additional resources from their surrounding environments since, given the limited amount of gas with which they initially formed, they would quickly burn through their entire supply by making stars.  Nonetheless, the intergalactic debris now seen within major galaxies such as the Milky Way are minor additions to already mature galaxies. Dwarf galaxies are analogous to interplanetary asteroids and meteoroids that continually impact Earth long after the bulk of our planet formed 4.6 billion years ago (Gya); the current terrestrial infall rate of ~40 kton/y, or an accumulated amount roughly equaling $2 \times 10^{17}$ kg over 4.6 Gy, is negligible compared to the mature Earth totaling $6 \times 10^{24}$ kg. Geologists do not consider our planet to have been forming throughout the past many billion years, rather that the bulk of Earth originated 4.6 Gya and has grown in small ways ever since. Likewise, most Milky Way development is now over, if not yet entirely completed, as building-block acquisitions continue to add <<1% of its total mass per encounter—much of it providing fuel for continued galaxy evolution as the assimilated galaxies, regardless of their small relative masses, bring in new stars, gas, and dark matter that occasionally trigger waves of star formation.

*4.1.2. Energy Rate Density for the Milky Way.* Our Galaxy today displays a 2-4-arm spiral geometry, probably with a linear bar through its center and visually measuring $~10^5$ light-years across a differentially rotating, circular disk of thickness $~10^3$ light-years. The entire system has been observationally estimated to contain $~10^{11}$ stars, of which our Sun is one of the great majority within the disk and $~2.6 \times 10^4$ light-years from its center. Visual inspection of stars and radio observation of nebulae show that our Galaxy's rotation remains nearly constant to a radial distance of at least $5 \times 10^4$ light-years, implying that the mass of the system within this radius is $~2 \times 10^{11}$ $M_\odot$, an extent delineated by its spiral arms comprising stars as well as much low-density interstellar matter. The integrated luminosity, L, or net energy flow in the Galaxy, measured at all wavelengths across the electromagnetic spectrum and including contributions from interstellar gas and dust, cosmic rays, and magnetic fields, as well as stars, is $~3 \times 10^{10}$ $L_\odot$ (or $~10^{37}$ W, where $L_\odot \approx 4 \times 10^{33}$ erg/s) within $5 \times 10^4$ light-years and very low surface brightness (if any luminosity at all) beyond [80]. Thus, *prima facie*, for the Milky Way, the energy rate density equals the inverse of its standard mass-to-light ratio: $(M/L)^{-1} \approx (7\ M_\odot/L_\odot)^{-1} = \Phi_m \approx 0.3$ erg/s/g.

The above estimates for M and thus for $\Phi_m$ do not include dark matter, an enigmatic ingredient of the cosmos that currently plagues much of modern astrophysics. If gravity binds our Galaxy, then such dark matter, which is probably mostly non-baryonic in nature, is needed to keep it from rotational dispersal; angular velocities of interstellar clouds in the Galaxy's extremities remain high far ($~10^5$ light-years) from the galactic center, the implication being that this huge physical system is even bigger and more massive, containing at least as much dark matter as luminous matter. Observations imply a diffuse spherical halo at least 10 times larger diameter ($~10^6$ light-years) than the visible disk [81], thus a Galaxy several times as massive as that given above (*i.e.*, $~10^{12}$ $M_\odot$), and a consequent value of $\Phi_m$ equal to at most a third of that derived above, or $~0.1$ erg/s/g. Order-of-magnitude lower values of $\Phi_m$ typically characterize the dwarf galaxies, whose luminosities are dim and masses dominated by dark matter, especially the eerie "dark galaxies" [82].

Here, we are concerned neither with the composition of the dark matter (the leading contenders for which are faint, massive compact halo objects [MACHOs] and invisible, weakly interacting elementary particles [WIMPs]), nor with the ongoing puzzle that this peculiar substance has so far eluded observational detection at any wavelength. Suffice it to say that an invisible halo apparently engulfs the inner domain of stars, gas, and dust once thought to represent the full extent of our Galaxy, and that the dark matter has much M yet little L, which then affect estimates of $\Phi_m$, hence presumably system complexity. We are in this paper not concerned about galaxies generally as much as the one we inhabit and of principal interest to big historians. By contrast to our Milky Way, the full range of values of $\Phi_m$ for all galaxies typically extends over an order of magnitude less for dwarf galaxies that usually harbor anomalously large amounts of dark matter, and perhaps two orders of magnitude more for active galaxies that are rare (~$10^{-4}$ of all galaxies) and whose emissions are beamed toward us during brief (<$10^6$-y) periods, making their abnormal flaring unrepresentative of such galaxies on average (cf., [17] for a fuller discussion of galaxies in general). All galaxies—whether normal, dwarf, or active—inhabit the lowermost part of Figure 2.

Figure 3 numerically summarizes the above discussion, plotting estimates of $\Phi_m$ for our evolving Milky Way, dating back to its origin ~12 Gya. This graph does not show sharp spikes of increased $\Phi_m$ that might have occurred during relatively brief (ten-to-hundred-My) episodes of enhanced star formation caused by significant (though unlikely major) encounters with neighboring dwarf galaxies—events that would have increased both M and especially L, thus potentially yet temporarily raising $\Phi_m$ by a few factors during the Galaxy's long mature phase. The only known flaring of its mostly dormant supermassive black hole, Sgr A$^*$, extends back only a few centuries when two events probably raised our Galaxy's L by less than a few percent and for less than a decade [83]. Astronomers have no reliable way to reconstruct the more distant past when star-bursts might have briefly, though dramatically, enhanced $\Phi_m$ during our Galaxy's evolution.

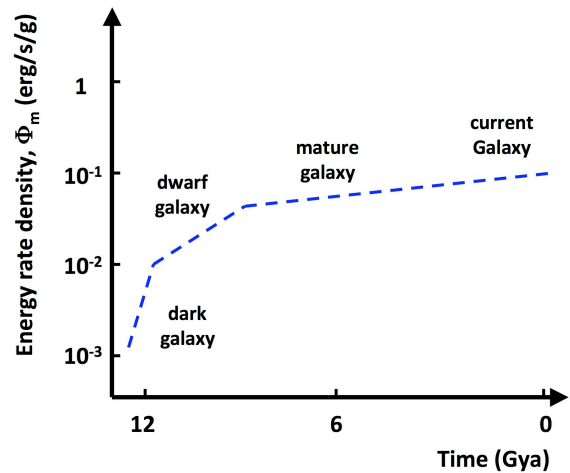

FIGURE 3: The growing complexity of the Milky Way Galaxy, expressed in terms of $\Phi_m$, is shown here rising slightly over its ~12 Gy existence to date during the physical-evolutionary phase of cosmic evolution. According to the hierarchical theory of galaxy construction, dwarf galaxies and pregalactic clumps of gas merged relatively rapidly in the earlier, denser Universe, such that within several billion years after the big bang our Galaxy had matured to nearly its present size and scale. The value of $\Phi_m$ for the Galaxy has continued rising ever since and will likely continue doing so, though only slightly, slowly, and episodically, as more galaxies (mostly dwarfs) collide and merge with our parent Galaxy.

*4.1.3. Galaxy Complexity.* We might expect that normal galaxies like the Milky Way would have values of $\Phi_m$ comparable to that of normal stars largely because, when examined in bulk, galaxies visually seem hardly more than gargantuan collections of stars. Yet galaxies contain much dark matter whereas stars do not. Since $\Phi_m$ is, effectively, an energy *density*, this quantity scales inversely as the mass of the entire galaxy housing those stars. As a result, galaxies typically have $\Phi_m$ values (0.01-50 erg/s/g) smaller than for most stars (2-1000 erg/s/g, see Sec 4.2), yet some overlap exists among the most active galaxies and the dimmest dwarf stars. Such overlaps in $\Phi_m$ should not surprise us, much as is the case sometimes for plants and animals or for society and technology (*cf.*, Sec. 4.5-4.7 below for life and civilization); outliers, exceptions, and overlaps, though rare, are occasionally evident among ordered systems in an otherwise chaotic Universe.

Since the onset of galaxies, in the main and in bulk, preceded most stars, and since galaxy values of $\Phi_m$ are typically less than those for stars, does that imply that galaxies are simpler than stars? And what about the common assertion that all life-forms are more complex than any star or galaxy (as stated earlier)? Life's inherent complexity stems not only from knowing that more data are needed to describe any living thing, but also that life manifests superior function as well as intricate structure; life-forms additionally and without exception do have larger $\Phi_m$ values as discussed below. As a general proposition, physical systems display less structural and functional complication, thus are likely simpler (although I formerly thought the opposite and once stated in print that galaxies are complex objects [84], but I now realize that by claiming that our Galaxy resembles a "galactic ecosystem . . . as complex as that of life in a tide pool or a tropical forest," I was parsing mere words to describe a subjective impression). In fact, galaxies *are* complex systems, yet their degree of complexity is evidently less than virtually any life-form and probably less than most stars as well.

That galaxies are simpler than expected by glancing at them is also not surprising from a systems perspective. Once their whole being is examined globally within their extended cosmic environments, galaxies are recognized to contain hardly more than $10^{9-12}$ relatively unordered stars. Ellipticals are the epitome of chaotically swarming stars; even spirals are ragged and misshapen when explored at high resolution—the disordered traces of a violent past. The many ongoing collisions, mergers, and acquisitions experienced by galaxies likely prevent them from becoming too complex. When they do collide the result is a mess, not some new order, much as when trains crash creating a wreck of simplified debris rather than a more ordered train. Sweeping spiral arms adorning some galaxies, as well as their cores, bulges, disks, and halos, are unlikely more complex than the many different components of stars—core, convection zone, photosphere, corona, as well as irregular spots and flares on stellar surfaces—indeed stars too are considered relatively simple based on their $\Phi_m$ measures (1-$10^3$ erg/s/g; see Sec. 4.2). All such physical systems are comparatively simple, at least in contrast with more complex, biological and cultural, systems that originated and evolved later in time.

Furthermore, the hierarchical model of galaxy formation, which holds that major galaxies are haphazardly assembled via many mergers of smaller pieces, implies that the properties of individual galaxies ought to be characterized by six independent parameters, specifically mass, size, spin, age, gas content, and their surrounding environment. But observational surveys of a wide variety of normal galaxies suggest that all these parameters are correlated with each other, and that in reality galaxy morphology may be dominantly regulated by a single such parameter—namely, their current mass [85, 86].

This does not mean that galaxy evolution is driven solely by gravity forces and energy flows resulting from conversion of gravitational potential energy, which can be readily modeled in coarse-grain N-body simulations. A suite of convoluted "gastrophysical" processes at local and regional levels within galaxies, including cooling and accretion of interstellar gas, transformation of that gas into stars, as well as feedback of energy and momentum from stars back into the gas, all comprise fine-grain, nature-nurture bookkeeping too disordered to currently simulate [87]. The formation, development, and evolution of galaxies, as minimally understood today from observations of different objects of different ages in different places, does display, *en masse*, simplicity transforming into complexity—the utter simplicity of the early primordial Universe giving way naturally to one in which matter is clumped, structured, and ordered. But complexity is a relative word and degrees of complexity are important; some organized matter that emerged after the onset of galaxies is even more complex, and hierarchically so—and that is what the term $\Phi_m$ seeks to quantify as a uniform, consistent, and general complexity metric for all ordered systems in Nature.

*4.1.4. Milky Way Summary.* Galaxies of all types, including those of dwarf, normal, and active status, have derived $\Phi_m$ values that are among the lowest of known organized systems—typically in the range 0.01 (dwarf types) - 50 (most active types) erg/s/g, with most normal galaxies displaying plus or minus a few factors times 0.1 erg/s/g. In the specific case of our Milky Way Galaxy, its $\Phi_m$ value rose while gradually developing:

- from protogalactic blobs >12 Gya ($\Phi_m \approx 10^{-3}$ erg/s/g)
- to widespread dwarf galaxies (~$10^{-2}$)
- to mature, normal status ~10 Gya (~0.05)
- to our Galaxy's current state (~0.1).

Although of lesser complexity and longer duration, the Milky Way is nearly as metabolic and adaptive as any life-form—transacting energy while forming new stars, cannibalizing dwarf galaxies, and dissolving older components, all the while adjusting its limited structure and function for greater preservation in response to environmental changes. By the quantitative complexity metric promoted here—energy rate density—galaxies are then judged, despite their oft-claimed majestic splendor, to be not overly complex compared to many other forms of organized matter—indeed unequivocally simpler than elaborately structured and exquisitely functioning life-forms.

*4.2. Our Sun*

*4.2.1. The Sun Today.* Our Sun is a typical G2-type star having a current luminosity $L_\odot \approx 4 \times 10^{33}$ erg/s (actually $3.84 \times 10^{33}$) and a mass $M_\odot \approx 2 \times 10^{33}$ g (actually $1.99 \times 10^{33}$), making $\Phi_m \approx 2$ erg/s/g today (more accuracy is unwarranted). This is the average rate of the Sun's energy release per unit mass of cosmic baryons, which fuse ~10% of their hydrogen (H) in 1 Hubble time (10 Gy). This energy effectively flows *through* the star, as gravitational potential energy during star formation converts into radiation released by the mature star. Specifically, the initial gravitational energy first changed into thermal energy to heat the interior, thence ignited nuclear energy in fusion reactions within the core, converted that energy to lower frequencies in a churning convection zone, and finally launched it as (mostly) visible electromagnetic energy from the mature star's surface. Such a star utilizes high-grade (undispersed) energy in the form of gravitational and nuclear events to build greater internal organization, but only at the expense of its surrounding environment; the star emits low-grade light, which, by comparison, is highly disorganized energy scattered into wider domains well beyond its internal structure.

Perspective is crucial, however. In the case of our Sun, ~8 minutes after emitting its light, life on Earth makes use of those dispersed photons, which though low-grade relative to the Sun's core are very much high-grade relative to the even lower-grade, infrared radiation that is, in turn, then re-emitted by Earth.  What is waste from one process (outflow from the Sun) can be a highly valued energy input for another (photosynthesis on Earth), as noted below in Sec. 4.4.

The cherished principles of thermodynamics remain intact. All agrees with the 2$^{nd}$ law of thermodynamics, which demands that entropy, or disorder, increases overall in any event.  The Sun's external environment is regularly disordered, all the while order emerges, naturally and of its own accord, within the stellar system per se—and eventually, indeed more so, within our planetary system that harbors life, intelligence, and society, all again as discussed in subsequent sections below.

*4.2.2. Evolution of the Sun.*  Once the young Sun entered the main sequence of normal stars and ignited H→He fusion, it remains hydrostatically balanced for ~11 Gy; its values of L and surface temperature $T_s$ change little.  Still, it is instructive to track those small changes, for they show that $\Phi_m$ does increase throughout the Sun's long lifetime, even in its relatively stable main-sequence phase.

Both theoretical inference and observational evidence reveal that our Sun currently increases its L at the rate of ~1% per $10^8$ y.  This occurs because, as the Sun fuses H→He within a central zone where the core temperature $T_c \geq 10^7$ K, the He ash accumulates and contracts, albeit slightly; much like a negative-feedback thermostat, the star continually adapts by readjusting its balance between inward gravity and outward pressure.  And as that ashen core "settles," it heats yet more to again rebalance against gravity, in the process fusing additional H within an expanding $10^7$-K shell overlying the core, thereby raising its energy production rate, though again only slightly—and very slowly.

This is the so-called "faint-Sun paradox" because life would have had to originate several Gya when Earth was unlikely heated enough to keep $H_2O$ liquefied since the Sun must have been dimmer than now when it first formed ~5 Gya.  The young Sun would also then have been somewhat more massive since it regularly loses mass via its solar wind, in fact it likely suffered an even greater mass loss during its youthful T-Tauri phase when its escaping wind likely resembled more of a gale while clearing the early Solar System of formative debris.  Although the Sun's early mass-loss rate is unknown, it was probably a small fraction of the star per se; today the Sun loses ~$2 \times 10^6$ metric tons of particulate matter per second (*i.e.*, $3 \times 10^{-14}$ $M_\odot$/y) and another $4.3 \times 10^6$ tons/s in equivalent radiation (*i.e.*, ~$6 \times 10^8$ tons/s of H converted to He at a nuclear efficiency of 0.71%), but that loss hardly affects the Sun as a star, diminishing its total mass by <<0.1% to date.  Computer models [88] imply that ~5 Gya the Sun was about half as luminous yet virtually as massive, making its L value at the time ~$2 \times 10^{33}$ erg/s and its $\Phi_m$ value early on ~1 erg/s/g.  Thus, over the past 5 Gy, $\Phi_m$ for the Sun has roughly doubled, and during the next 6 Gy will nearly double again by the time its central H fusion ends.

When the Sun does begin to swell toward red-giant status in ~6 Gy, it will experience a significant increase in $\Phi_m$ while evolving and complexifying more dramatically.  Post-main-sequence evolutionary changes accelerate in every way: Its L will increase substantially, its color will change noticeably, its internal gradients will grow greatly, and its value of $\Phi_m$ will rise much more rapidly than in its first 11 Gy. What follows are some numerical details of this evolutionary scenario, averaged over many models, noting that until nearly the star's demise M remains practically constant all the while L and therefore $\Phi_m$ increase [89, 90].

In ~6.2 Gy, the Sun's extremities will expand while exhausting H gas at its core, yet still fusing it within the surrounding layers. Its L will first become nearly twice larger (in addition to its already main-sequence doubled value of L today), making then $L_\odot \approx 10^{34}$ erg/s—the result of a bloated object fluxing its energy through a larger surface area as our future Sun becomes an elderly subgiant star. By then, its energy output will have increased because its core $T_c$ will have risen with the continued conversion of ever-more gravitational to thermal energy; He ash accumulating in the core will contract substantially, thus producing more heat, which once again stabilizes the star against collapse. By contrast, its surface $T_s$ will then have decreased as with any distended object from ~6000 K to ~4500 K, making its previous (as current) external color of yellow more orange. At this point, the star will have become a convoluted object—its envelope expanded past the size of Mercury's orbit while receding into interplanetary space and its core contracted to the size of Earth while approaching the quantum state of electron degeneracy. As its He-ashen core then continues compacting under the relentless pull of gravity, its $T_c$ will approach the $10^8$ K needed to fuse He, all the while its $T_s$ will have lowered further to ~4000 K and its surface reddened as the aged star inflates further.

Additional complications will become manifest since, although H→He fusion occurs throughout the more voluminous intermediate layers, that process will have switched from the simpler proton-proton cycle to the more elaborate CNO cycle (wherein those heavy nuclei, especially C, act as nuclear catalysts) mainly because the overlying layers will then be heated to higher T from the even hotter underlying core. Eventually, ~0.7 Gy after leaving the main sequence and following an extremely short period of unstable, explosive He fusion when it first ignites (or "flashes" ferociously for a few hours according to computer models), the star will attain a more stable state while it fuses He→C and displays $L \approx 50 L_\odot$, but only for ~$10^8$ y more—the classic late stage of a red-giant star near "death" [71].

Throughout this period of post-main-sequence evolution, the Sun's internal thermal, density, and elemental gradients will have markedly steepened; its mass will have decreased to ~0.8 $M_\odot$ owing to strong winds and serious mass-loss caused by its larger size (~100 $R_\odot$) and reduced surface gravity; and its core, once laden with mostly H fusing into He will have become mostly He fusing into C, all of which guarantees a more differentiated internal constitution—a clear sign of an evolved physical system that has become decidedly more complex, as are all red-giant stars.

Ultimately and for a much shorter period of time (<10 My) as He is consumed and C accumulates in its core, the elderly Sun will likely swell still more and lose more M while transitioning deeper into the giant domain, where its values of L and hence $\Phi_m$ probably increase by roughly another order of magnitude. Multiple shells of H and He will then fuse internally, but its total mass is likely too small to allow its core to reach $6\times10^8$ K needed to fuse C→O, thus its central fires will extinguish without synthesizing heavier nuclei beyond token amounts of O. While nearing its end fate, the Sun's constitution will have become more complicated than when it first began fusing as a homogeneous sphere of mostly H gas ~5 Gya. The future Sun will be unable to survive these changing conditions. It is destined for deletion from—that is, will be physically selected out of—the local population of stars.

*4.2.3. Energy Rate Density for the Sun.* The escalating complexity described here for a 1-$M_\odot$ star is well reflected in its increased $\Phi_m$ values throughout its stellar evolutionary journey—much as expected for any open, non-equilibrated system both evolving and complexifying. The Sun, in particular, has, and will have, increased its $\Phi_m$ values throughout its lifetime while repeatedly adapting (*i.e.*, adjusting) to its environmental circumstances. Figure 4 graphically summarizes these principal changes.

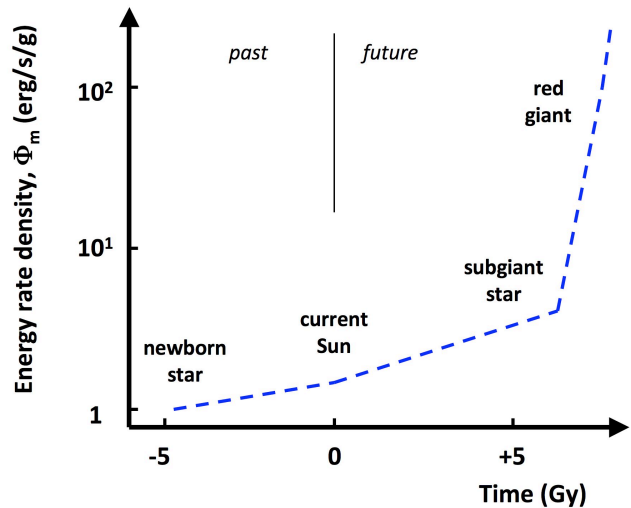

FIGURE 4: The value of $\Phi_m$ for the Sun increases gradually while fusing H→He throughout >95% of its total ~12-Gy lifetime (on the left side of the vertical line to the present and on the right side into the future). Even while on the main sequence for ~11 Gy, the Sun approximately quadruples its luminosity and hence its energy rate density while steadily, yet very slowly, growing internally more complex. Only toward the end of its tenure as a nuclear-burning star does the Sun's core contract enough to trigger He→C fusion, accelerate its internal organization, and cause a rapid rise in $\Phi_m$ by about an order of magnitude.

Rising $\Phi_m$ well characterizes the Sun as it becomes more structurally complex while physically evolving—but only while fusing as a genuine star. Its ultimate destiny is two-fold: a slowly receding outer envelope that gradually disorders by dispersing into the surrounding interstellar medium, and a small, dense, hot core remnant whose C embers glow solely due to its stored heat. These latter, white-dwarf stars are not stars per se (in contrast to red-giant stars that really are stars while still fusing nuclei); there is actually nothing stellar about a white dwarf since no nuclear fusion occurs within such a relatively homogeneous sphere of C that is supported only by a sea of electrons obeying the Pauli exclusion principle. Such an end-state for the Sun is not very complex—and not very surprising either, since such a dead star, as for any declining object, animate or inanimate, has a decreasing $\Phi_m$ value and thus an energy flow well below optimum.

More generally for all stars (*cf.*, [16] for a fuller discussion of complexity changes within stars that are more and less massive than the Sun), stellar interiors undergo cycles of nuclear fusion that foster greater thermal and chemical gradients, resulting in increasingly differentiated layers of heavy elements within highly evolved stars. Our Sun is the product of many such cycles. We ourselves are another. Without the elements synthesized in the hearts of stars, neither Earth nor the life it shelters would exist. Low-mass stars are responsible for most of the C, N, and O that make life on Earth possible; high-mass stars produce the Fe and Si that comprise the bulk of planet Earth itself, as well as the heavier elements on which much of our technology is based.

Growing complexity can, therefore, serve as an indicator of stellar aging—akin to developmental stages of immaturity, adulthood, and senescence for organisms [91]—while their interiors sustain fusion, thereby causing them to change in size, color, brightness, and composition while passing from "birth" to "maturity" to "death." (Even later in this paper when discussing life, biological evolution equates with the evolution of developmental life-cycles as well as the more common adaptation/selection process of generational neo-Darwinism.) In addition, stellar complexity also rises during even longer times—akin to more familiar evolutionary processes such as the growth of diversity in ecosystems [92, 93]—as stars change over multiple generations in space [16]. Such changes very slowly alter the constitution of every star, and the Sun is no exception. At least as regards energy flow, complexifying structure, and growing functionality while experiencing change, adaptation, and selection, stars have much in common with life.

*4.2.4. Sun Summary.* On and on, the cycles roil; build up, break down, change. Stars adjust their states while evolving during one or more generations, their energy flows (per unit mass) and their $\Phi_m$ values rising while they complexify with time. In the case of the Sun:
- from early protostar ~5 Gya ($\Phi_m \approx 1$ erg/s/g)

- to the main-sequence Sun currently (~2)
- to subgiant status ~6 Gy in the future (~4)
- to aged red-giant near termination (~$10^2$)
- to black dwarf status (0 erg/s/g) as its nuclear fires cease, its envelope dissipates, its core shrivels and cools, and its whole being fades to equilibrated blackness—but not for a long, long time greater than the current age of the Universe.

*4.3. Planet Earth*

*4.3.1. Earth Internally.* Much of Earth's original organization derives from energy gained from accretion of mostly homogeneous, proto-planetary matter in the early solar nebula. Conversion of gravitational potential energy into thermal energy, supplemented by radioactive heating, created energy flows that helped promote Earth's geological complexity, from center to surface. In particular, during Earth's formative stage ~4.6 Gya when it experienced much of its gross ordering into core, mantle, and crust, its internal value of $\Phi_m$ was much larger than now. This is not surprising since almost all of our planet's early heating, melting, and differentiating occurred before the oldest known rocks formed ~4.2 Gya. Its initial energy rate density then characterized the thermal and chemical layering within the early, naked Earth (minus a primordial atmosphere that had escaped, an ocean that was only starting to condense, and a biosphere that did not yet exist); remnants of the internal bulk of our planet are what geologists explore and model today.

Unlike gaseous stars that continue increasing their thermal and chemical gradients via physical evolution often for billions of years after their origin, rocky planets complexify mostly in their formative stages while accreting much of their material in <$10^8$ y, after which internal evolutionary events of a geological nature comparatively subside. It is during these earliest years that planets, at least as regards their bulk interior composition, experience the largest *internal* flows of energy in their history. Note again that this subsection does not address Earth's external atmosphere, ocean, and biosphere that later developed on our planet—and for which $\Phi_m$ would eventually rise (see below).

The current value of $\Phi_m$ for the entire rocky body of Earth per se is negligible in the larger scheme of cosmic evolution since the bulk of our planet's interior is not now further complexifying appreciably. Earth's internal energy flow, mostly in the form of stored heat upwelling from within, derives from three sources: gravitational contraction of its formative matter and the sinking of mass concentrations of heavy elements (notably Fe and Ni) toward the core while differentiating, accretion of additional matter during a period of heavy meteoritic bombardment up to ~3.8 Gya, and lingering radioactive decay of heavy unstable nuclei (like Al and K) originally acquired from the supernova debris of nearby massive stars. All these events together, today and long past their peak, yield a small energy outflow at Earth's surface, measured and globally averaged to be ~63 erg/cm$^2$/s [94]. When integrated over the entire surface of our planet's globe, this equates to an effective (geothermal) luminosity of ~3.2x10$^{20}$ erg/s (or 32 TW).

Since Earth's mass totals ~6x10$^{27}$ g, then $\Phi_m \approx$ 5x10$^{-8}$ erg/s/g for our planet's interior today—an energy rate density consistent with a minimally ordered yet relatively unchanging physical object (globally considered), much like an already formed, mostly solidified, and largely dormant crystalline rock having $\Phi_m \approx$ 0—which, by the way, much of Earth internally is. Even this small heat flow, however, can affect planetary evolution at the surface locally, while driving events with implications for life; tectonic activity represented by recent mountain-building or volcanism such as the Alps or Hawaii have current $\Phi_m$ values typically twice that of geologically old and inactive areas such as the pre-Cambrian shields. Mid-oceanic trenches are sites of greatest radiogenic heat flow at or near the surface of Earth today, reaching values of ~150 erg/cm$^2$/s, and sometimes double that in especially active underwater vents. Rich mineral deposits, found geologically in Earth's crust where condensation of hot fluids are driven by temperature gradients, display substantial, yet local, internal energy flows, hence abiotic complexity, as do hurricanes, tornadoes, and other meteorological phenomena driven externally by solar energy [95, 96]. However, this paper mainly addresses our planet globally and historically, leaving aside for now smaller-scale regional effects.

Earlier in Earth's history, when our planet was changing more rapidly during its first ~1 Gy—developing, settling, heating, differentiating—its value of $\Phi_m$ would have been much larger. Taking a surface temperature, T ≈ 1800 K [97] as an average value of a "magma ocean" during its initial 0.5 Gy, and knowing that energy flux through a surface area scales as $\sigma T^4$ (where $\sigma$ is the Stefan-Boltzmann constant = 5.7x10$^{-5}$ erg/cm$^2$/K$^4$/s), we estimate that in Earth's formative years its energy rate density would have been enhanced by (1800/256)$^4$, making $\Phi_m$ *then* several orders of magnitude larger than now. (A surface temperature of 256 K is used in this calculation, not 288 K as is the case today, since the former is the "thermally balanced temperature" when the incoming solar energy absorbed equaled the outgoing terrestrial heat emitted for our early naked planet, whereas 288 K is Earth's "enhanced greenhouse temperature" boosted in more recent times by the thickening of our planet's atmosphere.)

*4.3.2. Primordial Earth.* Earth's original value of $\Phi_m$ can be estimated by appealing to the conservation of energy, here the 1$^{st}$ law of thermodynamics applied to a massive body governed by the gravitational constant G (6.7x10$^{-8}$ cm$^3$/g/s$^2$). Setting the gravitational potential energy of a gas cloud of mass, M, that infalls to form a ball of radius, r, during a time interval, t, equal to the accreted energy gained and partly radiated away while converting that potential energy into kinetic energy, which in turn causes a rise in surface temperature, we find:

$$\tfrac{1}{2}(GM^2/r) = 4\pi r^2 t\sigma T^4 \ .$$

The right side of this equation equals the total energy budget of the proto-planetary blob, namely the product of luminosity (L) and duration (t). The fraction $\tfrac{1}{2}$ results from the commonly accepted Virial Theorem, which specifies that half of the newly gained energy of any contracting mass radiates away, lest the formative process halt as heat rises to compete with gravity; that escaped part of the energy budget does not participate in formative ordering. The result for early Earth was significant heating, indeed melting, mostly via gravitational accretion and later by the decay of radionuclides; however, none of the most abundant radioactive elements, including U and Th, have half-lives short enough to have participated in much of this early heat pulse, thus they are neglected in this approximation.

Accordingly, an estimated value of $\Phi_m = GM/2rt \approx 10$ erg/s/g characterizes the young Earth, an energy rate density generally larger than the less-ordered Sun (see Sec. 4.2) yet smaller than Earth's subsequently more-ordered biosphere (see Sec. 4.4), much as expected if energy rate density is a complexity metric for organized systems experiencing cosmic evolution. With $t \approx 10^{3-4}$ y, we also find $T \approx 3000$ K, a not unreasonable temperature to which ancient Earth might well have been heated during its accretional stage [98], in fact much less than the ~60,000 K to which the assembled rocky planet would have been heated had all the produced energy been stored internally. The time scale for terminal accretion, that is, the total duration needed to sweep clean the primitive Solar System and to form each of the planets, is more like $10^{7-8}$ y, but the solar nebula cooled and its mineral grains condensed on the order of $10^4$ y. During this latter, shorter time interval the bulk of the planets likely emerged; otherwise, loose matter in the solar nebula would have been blown away by strong "T Tauri," bipolar solar winds [71]. By contrast, slower accretion over the course of millions of years would have allowed the newly gained heat to disperse, resulting in negligible influence on its internal temperature (typically a few hundred K) and thus an inability to melt rock (as opposed to merely heating it), causing minimal geochemical differentiation, if any—which we know from Earth's exploration is not what happened.

As calculated above for more rapid accretion, $T \approx 3000$ K was surely high enough to melt rock, thus helping (along with some short-term decays of radionuclides like Al) to order our planet's interior as the low-density materials (rich in Mg and Si) percolated toward the surface while the high-density materials (rich in Ni and Fe) sank toward the core—yet not such a high temperature as to make this analysis unrealistic. In turn, the long-lived radionuclides (U and Th) and the potential energy realized when huge globs of molten metal plunged radially downward would have further heated Earth's core enough to establish a robust magnetic field from the dynamo action of mostly spinning iron. The result is a planet that today is well differentiated, with moderate density and temperature gradients extending from core to surface: ~12 g/cm$^3$ - 3 g/cm$^3$ and ~6000 K - 290 K, respectively.

All these heating, fluxing, and ordering events occurred long ago on Earth. Currently, when averaged over our entire planetary globe, $\Phi_m$ internally is very much smaller (~10$^{-7}$ erg/s/g, as computed above), nor is there much ordering now occurring internally apart from a few "hot spots" that drive today's surface tectonic activity—and, of course, in the climasphere and biosphere, where much externally enhanced order is indeed evident, not from energy flowing outward from inside Earth but that flowing inward from outside, indeed from the Sun.

*4.3.3. Earth's Climasphere.* Planets are often appraised to be more complex than either stars or galaxies, thus it is not surprising that planetary values of $\Phi_m$ are also somewhat larger—at least for some parts of some planets at some time in their history. (That is why arguably more is known about the Sun than the Earth; stars are simpler systems.) Here we examine not our planet's whole globe, from its interior through its surface, since Earth is not now evolving much ~4.6 Gy after its origin. Rather, what is most pertinent in this analysis are those parts of our home that are still evolving robustly, still requiring energy to maintain (or regenerate) their structure and organization—indeed now fostering energy-rich and rapidly changing environments conducive to the emergence of even more complex systems, including animated life and cultured society.

Consider, for example, the amount of energy needed to power Earth's climasphere, which is the most highly ordered part of our planet today. The climasphere includes those parts of the lower atmosphere and upper ocean that absorb (and then re-emit) solar radiation, and which most affect turbulent meteorological phenomena capable of evaporating copious amounts of water as well as mechanically circulating air, water, wind, and waves. The total solar radiance intercepted by Earth is 1.8x10$^{24}$ erg/s, of which 69% penetrates the atmosphere (since Earth's global albedo is 0.31). This external power is several thousand times that currently present at Earth's surface from its warm interior. Photosynthesis is an inherently inefficient process (~0.1% overall; *cf.*, Sec. 4.4), so the great majority of incoming solar energy serves to heat the surface as well as drive atmospheric motions and ocean currents. Since our planet's air totals ~5x10$^{21}$ g (mainly the troposphere to a height of ~12 km, which contains >90% of the total atmospheric mass) and the mixed ocean layer engaged in weather (to depth of ~30 m) amounts to about double that mass, $\Phi_m$ for planet Earth today is roughly 75 erg/s/g.

Incidentally, the infrared (~10$^4$-nm wavelength) photons, re-emitted by Earth and equal in total energy to captured sunlight reaching the surface, are both greater in number and lower in energy (~20 times difference per photon) than the incoming sunlight of yellow-green (~520-nm) photons, thus contributing to the rise of entropy beyond Earth, even as Earth itself grows more complex and less entropic—again in accord with the 2$^{nd}$ law of thermodynamics.

To reiterate what has been computed here: The whole mass of planet Earth is not used in this mass-normalized, energy-rate calculation for the climasphere for two reasons. First, the heat flux generated internally by our planet is only a minute fraction of Earth's total energy budget today, thus can be neglected. Second, the incident solar radiation, which now dominates that budget, is deposited mainly into the external surface layers of Earth's atmosphere, upper ocean, and biosphere, from which it is then re-radiated into the dark night sky; solar energy does not flow through the interior of our planet. Only the mass of the climasphere is relevant in this particular $\Phi_m$ computation, for it is only in this air-ocean interface that solar radiation at Earth affects and maintains ordering on our planet today. Recent non-linear climate modeling confirms that energy rate density plays a dynamical role in horizontally stratifying the atmosphere and upper ocean, nurturing climaspheric complexity and offering confidence in the above analysis [99].

*4.3.4. Energy Rate Density for Earth.* Planetary systems generally, and Earth in particular, can be quantitatively analyzed in much the same way as for stellar and galactic systems above, especially regarding the rise of complexity and its hypothesized metric, $\Phi_m$. Energy flow, physical evolution, and system adjustment help us understand how typical planets are comparable to or slightly more complex than normal stars, and not least how on one such planet—the third body out from the Sun—conditions changed sufficiently for Nature to foster the emergence of even more highly organized biological life.

Figure 5 plots estimates of $\Phi_m$ as Earth's system dynamics evolved over the course of the past ~4.5 Gy, initially ordering and complexifying its mantle-core-crust interior, then later (including now) its ocean-atmosphere exterior. Soon after its formation, our planet had an internally driven value (surface through core) of ~10 erg/s/g, and an externally driven value that was negligible (since neither atmosphere nor ocean then existed). Later, the relative values of $\Phi_m$ reversed; internal energy flows weakened as rocky Earth cooled, while external energy flows strengthened within its gas-liquid climasphere because both the Sun's luminosity increased over time and our planet's ocean and atmosphere developed. Today, Earth's internally sourced value of $\Phi_m$ is insignificant compared to its externally sourced ~75 erg/s/g. Overall, combining internal and external contributions, Earth's total value of $\Phi_m$ rose somewhat during our planet's physical evolution as its atmosphere-ocean eventually formed and organized after the bulk of Earth had internally differentiated.

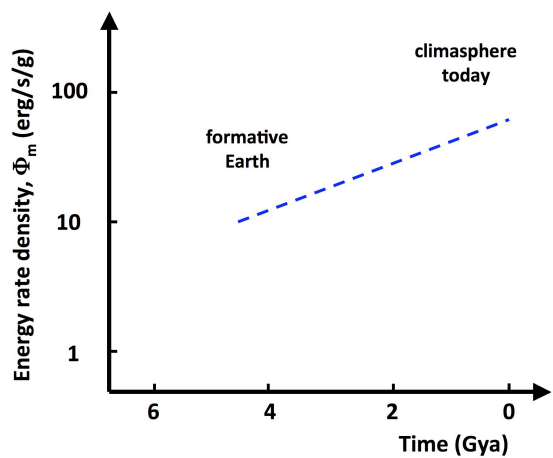

FIGURE 5: The value of $\Phi_m$ for Earth increased by roughly an order of magnitude over the course of our planet's ~4.5-Gy history as internal heat first structured our planet's solid interior and then external solar energy organized its gas-liquid climasphere. This graph pertains to the middle part of the master graph in Figure 2, much as expected for any relatively simple inanimate system complexifying during the physical-evolutionary phase of cosmic evolution.

*4.3.5. Earth Summary.* Qualitatively, our cosmic-evolutionary scenario seems to be holding as a scientific narrative that grants some appreciation for the rise in complexity as galaxies, stars, planets, and (below) life-forms emerged in turn. Quantitatively, despite some approximations of past events that are only partly understood, our energy-flow calculations compare and contrast reasonably well for known physical systems that have originated along the arrow of time. Since Earth's value of $\Phi_m$ today (~75 erg/s/g) exceeds that currently for the Sun (~2) as well as for the Milky Way (~0.1), Earth can indeed be reasonably judged a more complex system than either its parent star or parent galaxy, albeit not dramatically so. In sum, much of Earth's structural complexifying would have been managed by ancient, internal energy flows, which long ago rather quickly developed its organized layered stratification, core rotation, and mantle convection, after which little further ordering occurred except externally near the surface by means of current, external energy flows that sustain our biosphere today:

- from its formative stage ~4.5 Gya internally ($\Phi_m \approx 10$ ergs/s/g)
- through Earth's middle history, which is mostly unknown
- to its current climasphere externally (~75).

Comparative planetology helps to gauge order, flow, and complexity among planets generally by applying the same kind of thermodynamic analyses performed above for stars and galaxies. Our focus here has been on Earth since our own planet is of most interest to big historians. Exploring further along the arrow of time, we shall find that these very same diagnostic tools also provide useful complexity measures of life, society, and machines, as discussed in the next several sections.

Do note that the thermal gradients needed for energy to flow in Earth's biosphere could not be maintained without the Sun's conversion of gravitational and nuclear energies into radiation that emanates outward into unsaturable space. Were outer space ever to become saturated with radiation, all temperature gradients would vanish as equilibrium ensued, and life among many other ordered structures would cease to exist; this is essentially a version of Olber's paradox, a 19$^{th}$-century intellectual puzzle inquiring, in view of the myriad stars in the heavens, why the nighttime sky is not brightly aglow. That space is not now saturated (or the night sky fully illuminated) owes to the expansion of the Universe, thus bolstering the suggestion that the dynamical evolution of the cosmos is an essential condition for the order and maintenance of all organized things, including not only Earth, but also life itself. All the more reason to welcome life within our cosmic-evolutionary cosmology, for the observer in the small and the Universe in the large are not disconnected.

*4.4. Plants*

The most widespread, and probably most important, biological process occurring on Earth today is plant photosynthesis, which produces glucose ($C_6H_{12}O_6$) for system structure and adenosine triphosphate (ATP) that acquires, stores, and expresses solar energy throughout the floral world. This biochemical activity displays a general, yet robust correlation between the degree of complexity and biological evolution, especially since plants' energy budgets are well understood and their fossil records (aided by genetic clocks) extend far back in time.

By contrast, animals' most prominent process is respiration, whereby oxygen ($O_2$) converts consumed carbohydrates into the organics of tissue structure and synthesizes ATP in mitochondria, which can then release energy when needed for bodily activities. While it is often said that plants are producers and animals consumers, in fact both engage energy as an essential ingredient of life. We consider plants first and then examine animals in Sec. 4.5 below.

*4.4.1. Oxygen Buildup.* The roots of photosynthesis date back at least 3 Gy, when rocks of that age first trapped chemicals that facilitate this autotrophic process in plants, algae, and some bacteria (but not archaea). The earliest practitioners likely used $H_2S$ rather than $H_2O$, and some of them probably practiced chemosynthesis, a similar (yet heterotrophic) process that utilizes the chemical energy of inorganic compounds and is thus not dependent on solar energy. Still extant on Earth today, these primitive, non-oxygenic chemosynthesizers include some of the oldest known fossils.

Unicellular, aquatic, yet still prokaryotic cyanobacteria that do use $H_2O$ appeared later, starting ~2.7 Gya (late Archean, which formally ended 2.5 Gya), when traces of chlorophyll and oil biomarkers become evident in the geological record and when banded-iron formations (BIFs) are first seen in ancient sedimentary rocks as $O_2$ began combining in Earth's oceans with dissolved Fe (having upwelled from the interior via hydrothermal vents and from erosion of surface layers) to precipitate minerals such as hematite ($Fe_2O_3$) and magnetite ($Fe_3O_4$) that drop to the seafloor; BIFs differ from the FeO-rich "red beds" that came later with the widespread appearance of rusty red-rock strata beginning ~1.8 Gya. Thereafter (probably when the seas became nearly saturated with it), $O_2$ began accumulating in the atmosphere ~2.3 Gya, heralding the so-called Great Oxygenation Event, a gradual build-up of free $O_2$ that was likely also aided by a decrease in atmospheric $CH_4$, which until then was a major (primordial) atmospheric gas [100]. Rock weathering studies imply that $O_2$ accumulated in the air only slowly, reaching 10% abundance ~0.8 Gya and current concentrations (~21%) only ~0.3 Gya.

Early oxygenic, photosynthesizing plants were then, as now for blue-green algae (cyanobacteria), simple organisms, having genomes of merely (1-9)x$10^6$ nucleotide bases (*i.e.*, ~1000 times less than for humans). Even so, the enriched air eventually fostered the emergence of somewhat more complex life-forms, most notably unicellular, eukaryotic protists; fossils imply that this momentous event occurred ~1.7 Gya by means of a mutually beneficial symbiotic relationship when a small, anerobic, prokaryotic cell engulfed a free-living, respiring bacterium, thereby initiating evolution of most, and perhaps all, eukaryotes [101]—and it probably happened when minute bacteria (today's powerhouse mitochondria) within the larger, fused cell discovered how to liberate more energy from food plus $O_2$ and thus could afford to have more genes [102]. Apparently, energy use was strategically at the heart of this singular evolutionary step in the history of life (eclipsed only perhaps by the origin of life itself), as with all major milestones in cosmic evolution. Although prokaryotes have remained unicellular without complexifying, in fact dominating life on Earth for ~2Gy, it was this bioenergetic innovation that likely permitted eukaryotes to emerge, evolve greater complexity, and eventually foster multicellularity. In turn, yet only as recently as ~550 Mya, one of those protists likely experienced a second symbiosis (or "serial endosymbiosis") with a cyanobacterium. The result was a chloroplast—the specialized organelle comprising leaf cells rich in chlorophyll molecules where photosynthesis occurs in all plants—a key ancestral feature of every modern plant, which include such familiar organisms as trees, herbs, bushes, grasses, vines, ferns, and mosses.

*4.4.2. Plant Evolution.* In post-Cambrian times, plants likely evolved from protists, notably green algae having physical structures and metabolic functions closely resembling those of today's photosynthesizing organisms. The story of plant evolution is reasonably well documented [103-105]:

Calcified fossils of multicellular green algae, which are freshwater organisms that are also capable of surviving on land, date back at least to the mid-Ordovician (~470 Mya), and possibly even earlier in the late-Cambrian (~500 Mya); the oldest specimens found represent several genera and thus were already diversified.  Mosses were among the first full-time inhabitants of the land, taking up residency early in the Paleozoic (~450 Mya).  Vascular plants, having internal plumbing with leaves, stems, and roots akin to those of modern plants, originated during the late Silurian (~420 Mya, some fossils preserved) and by the mid-Devonian (~380 Mya, many fossil examples) had greatly multiplied and diversified while spreading into copious environments, thereby creating the first forests.  Primitive seed plants emerged near the start of the Carboniferous (~360 Mya), though most such species perished during the Permian-Triassic mass extinction (~252 Mya).  These earliest seed plants were the gymnosperms, whose "naked" seeds are not enclosed in protective structures and whose modern types include evergreen trees such as conifers and pinewoods.  The angiosperms, by contrast, comprising the flowering plants with sheltered seeds as with most grasses and deciduous trees, were the last major group of plants to appear, evolving rather suddenly from among the gymnosperms during the early Cretaceous (~125 Mya) and then rapidly diversifying ~30 My later.  Although the gymnosperms (~1000 species today) ruled life for at least 250 My, angiosperms (~350,000 species today) later crowded them out; ~90% of land plants are now angiosperms and a nearly continuous record of their fossils is preserved in rocks over the past 50 My.

   Not all these evolutionary strides likely happened fast and episodically because of dramatic environmental changes triggered by asteroid impacts or volcanic upheavals.  Many, and perhaps most, of these changes probably occurred gradually owing to a variety of environmental stresses, including drought, salinity, and cold.  A central hypothesis proffered here (*cf.*, Sec. 5.3) is that optimal use of energy played a significant role in these biological evolutionary steps as with all evolutionary advances.

*4.4.3. Photosynthesis Efficiency.*  Living systems generally require larger values of $\Phi_m$ than inanimate systems, not only to maintain their greater structural order in tissues and fiber but also to fuel their complex functions of growth, metabolism, and reproduction.  Plants, in particular and on average, need $1.7 \times 10^{11}$ ergs for each gram of photosynthesizing biomass, and they get it directly from the Sun.  *SeaWiFS* satellite sensing shows the global conversion of $CO_2$ to biomass is $\sim 2 \times 10^{17}$ g annually (*i.e.*, about twice 105 Gtons of C net primary production [106]), so Earth's entire biosphere uses energy at the rate of $\sim 10^{21}$ erg/s [107, 108].  This is ~0.1% of the total solar power reaching Earth's surface (~90 PW), therefore the electromagnetic energy of only ~1 in 1000 photons is converted into chemical energy of plants.  Even at that low efficiency of energy conversion, photosynthesis represents the world's largest battery; it stores huge quantities of energy both in living plants as well as dead plants ("fossil fuels") as coal, oil, and gas. Expressed in units of the complexity metric preferred in this paper, given that the total mass of the terrestrial biosphere (*i.e.*, living component only, >99% of it in the form of uncultivated land biomass and ~90% of that in forests) is ~$1.2 \times 10^{18}$ g (or ~teraton, an average from many researchers, not including any potential "deep hot biosphere" [109]), the value of $\Phi_m$ for the biogeochemical process of photosynthesis is, again globally averaged for the vast majority of Earth's plant life, ~900 erg/s/g.

It is often said that photosynthesis is a highly efficient process that is not understood, whereas in reality it is a very inefficient process that is rather well understood. Photosynthesis is limited by a wide range of variables, including light intensity, $CO_2$ abundance, $H_2O$ availability, environmental temperature ($T_e$), and leaf morphology, all of which interact in complicated ways; the process also has optimal ranges for each of these variables, such as a minimum $T_e$ below which and a maximum $T_e$ above which photosynthesis will not operate [110]. Photosynthesis is inherently inefficient for the complete metabolic process that converts sunlight into chemical energy stored in glucose molecules, *i.e.*, a ratio of output to input energies—not the higher rate (or effective absorptivity, which can reach as much as 65-90% depending on the species) of solar photons splitting $H_2O$ and releasing electrons. At the molecular level, the maximum quantum efficiency is ~28%. But only 45% of solar radiation is within the visible electromagnetic band (400-700 nm) where the light-harvesting pigment chlorophyll-a is active (trapping red and blue light, yet reflecting green), thereby reducing the actual molecular efficiency to only ~12%. Furthermore, ~$1/3$ of the absorbed energy is needed to power plant respiration, and ~$1/5$ of sunlight is typically blocked by overlying canopy, leaving only ~6.5% as the theoretical maximum efficiency of any plant [111].

Operationally then, photosynthesis suffers high losses, converting into chemical energy only ~0.1% of the incoming solar energy falling onto a field of uncultivated plant life [107]; this very low efficiency is actually due more to limited supplies of atmospheric $CO_2$ than lack of energy (usually because leaves' pores, depending on weather conditions, only partially open and thus deprive some plants from adequate supplies of $CO_2$). The value for $\Phi_m$ (900 erg/s/g) computed above, which is valid for the great majority of Earth's lower plant life, is sufficient to organize cellulose (the main carbohydrate polymer of plant tissue and fiber) for a field of wild plants, hence for the great bulk (>90%) of Earth's untended flora. And, as with the energetics of any complex system, energy re-radiated as waste heat fundamentally causes an entropy rise in the surroundings, thereby adding to the natural thermal balance of Earth's atmosphere in accord with thermodynamics' $2^{nd}$ law.

*4.4.4. Advanced Plants.* More organized fields of higher-order plants such as herbs and shrubs, and especially cultivated crops such as rice and wheat, can photosynthesize more than an order of magnitude more efficiently (1-2%) than the global average; their values of $\Phi_m$ are typically in the range of 3000-18,000 erg/s/g. Abundant deciduous trees have larger absorbing leaves that capitalize on the short, hot summers by photosynthesizing fast, yet their leaves die young compared to evergreen trees that achieve slower, steadier growth year-round; averaged annually, net productivity and efficiency of the two types of trees are comparable, 0.5-1%, implying $\Phi_m$ = 5000-10,000 erg/s/g.

Amongst the rarest of plants, the more advanced and complex $C_4$-type (that initially fix $CO_2$ around the key enzyme RuBisCO to make 4-carbon sugars, such as for maize, sorghum, millet, amaranth and sugarcane, but also including some of the worst weeds such as crabgrass) have photosynthetic efficiencies about twice (*i.e.*, 2-3.5%) that of the simpler, more widespread $C_3$-type plants (such as rice, wheat, barley, beans, potatoes, tomatoes, and sugar beets that have 3-carbon sugars). This is probably so because the specialized $C_4$ pathway—nonetheless practiced by ~7500 species of plants today, mostly grasses—uses less $H_2O$ and $CO_2$, employs greater nutrient uptake, and displays longer growth cycles, although both use the Calvin-cycle to facilitate $CO_2$ assimilation.

Empirical records imply that $C_4$ plants evolved from their $C_3$ ancestors only as recently as ~20 Mya (fossil dating) or ~30 Mya (genetic clock), in any case well after the Cretaceous-Tertiary geological boundary and even long after the appearance of the first $C_3$ grasses ~60 Mya. The $C_4$ pathway likely arose as a competitive advantage either while coping with high-temperature droughts or reduced $CO_2$ levels (atmospheric $CO_2$ levels did decline rapidly from ~1000 to ~500 ppm between 25 and 30 Mya), or while adapting to open, tree-less environments, and maybe for a combination of all these reasons—yet did so independently on at least 45 separate occasions and therefore along >45 separate lineages thereafter [112]. Only more recently did the $C_4$ photosynthetic upgrade cause grasses to transform the warm-climate subtropics, converting forests to grass-dominated savannahs between 3 and 8 Mya [113]. Very much more recently, these grasses were additionally subjected to cultural evolution as our ancestors during the past 10 ky sought to breed crop production for agricultural purposes by making photosynthesis yet more efficient. Some of these crop efficiencies will likely be bettered once again as ways of growing genetically modified crops become enhanced in today's technological society, such as current attempts to replace rice's inefficient $C_3$ pathway with the $C_4$ route found in maize and several other plant species able to produce a good deal more carbohydrates for a given energy/resource input—but this is mostly cultural, not biological evolution.

Cultivated plants do display higher values of $\Phi_m$, yet altogether produce <1% of the total yield of organic matter globally [114]. The most highly cultivated $C_4$ plants, such as maize and sugarcane that have been made more efficient (~2.5%) by advanced agricultural practices of recent times, probably cannot be fairly compared with fields of wild grasses and genetically unaltered trees and shrubs. Such well tended fields display higher energy rate densities, not only because enhanced values of $\Phi_m$ are consistent with increased metabolism of the more evolved tropical plants, but also because improved organization of fields produced by modern agricultural methods requires higher $\Phi_m$ values to maintain that organization—the latter reason reliant again on an energy contribution of a cultural, technological nature.

Independent evidence also suggests that energy use was likely a factor in the evolution of more advanced species of plants, especially the stunning diversification and rapid rise to ecological prominence of the angiosperms in the mid-to-late Cretaceous—an evolutionary event colloquially termed "Darwin's abominable mystery" (for there was nothing gradual about it). Angiosperms have higher growth rates and nutrient needs than gymnosperms; they sequester more N and P in their leaves, which then decompose quicker and thus, by positive feedback, create richer soil conditions for their own growth. Hence, the angiosperms probably utilized higher energy budgets than gymnosperms, allowing the former to out-compete the latter during one of the greatest terrestrial radiations in the history of life [115]. Furthermore and theoretically, hierarchies in energy density (if not energy rate density) have long been expected for organisms ascending the trophic ladder in ecosystems [59, 116]. By contrast, plants struggling under hot, arid conditions—such as the slow-growing succulents cacti and pineapple—photosynthesize mostly at night by means of a different process known as Crassulacean acid metabolism (CAM), which evolved to minimize $H_2O$ losses, and these are among the least efficient in the living world.

*4.4.5. Energy Rate Densities for Plants.* Figure 6 summarizes average values of $\Phi_m$ for a variety of members within several plant categories. As for galaxies, stars, and all inanimate systems, $\Phi_m$ values for animated life-forms range considerably, often over an order of magnitude or more—as here for gymnosperms (such as pine, fir and larch evergreen trees), angiosperms (such as oak and beech deciduous trees or wheat and tomato herbs), and tropical $C_4$ grasses (such as maize and sugarcane). Variations in $\Phi_m$ occur among plants because they do not equally absorb incoming sunlight and do not convert with equal efficiency harvested energy into biomass [117, 118].

The plotted values, relative to all plants in Earth's biosphere generally having an efficiency of 0.1% and $\Phi_m \approx 900$ erg/s/g, clearly display an increase over time. The flowering angiosperms (with their more specialized fiber-cell anatomy and more intricate reproductive system) are widely considered more botanically complex than the unprotected-seed gymnosperms [119]. Taken together, all the computed plant values of $\Phi_m$ *generally* agree with a central hypothesis of this review paper, namely that normalized energy flow, biological evolution, and increased complexity are reasonably well correlated. Although we have crossed over into the realm of living systems, energy rate density remains a potentially useful way to quantify the rise of complexity during biological evolution, much as done elsewhere in this paper for many other complex systems experiencing simpler physical evolution and more complex cultural evolution throughout Nature.

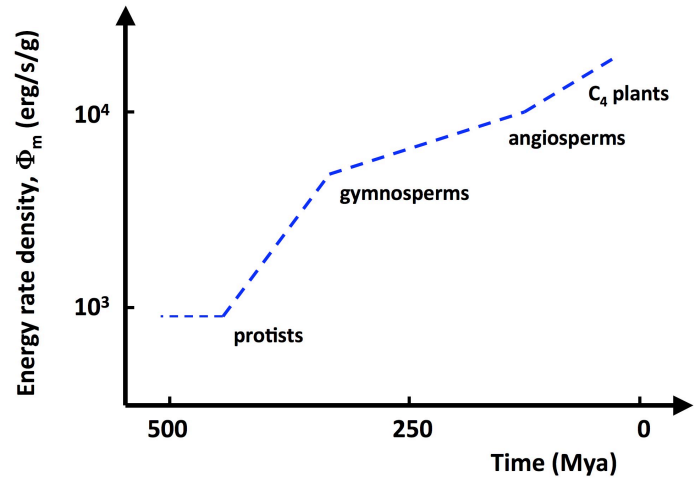

FIGURE 6: The complexity of plants, expressed in terms of $\Phi_m$, includes a range of increasingly ordered structures for a wide variety of photosynthesizing life-forms at various stages of the biological-evolutionary phase of cosmic evolution. Note how flowering angiosperms have higher energy rate densities than gymnosperms or protists, and, in turn, more organized, cultivated $C_4$ crops such as maize and sugarcane still higher values—yet all such plants on Earth have $\Phi_m$ values within a factor of ~20 of one another.

*4.4.6. Plant Summary.* Plants regularly exhibit intermediate values of $\Phi_m \approx 10^{3-4}$ erg/s/g—well higher than those for galaxies, stars, and planets, though lower than for animals, society, and machines (see below). Average values discussed here are typical of a variety of photosynthesizing plants found on Earth during post-Cambrian times—all of them nestled within the middle part of the cosmic-evolutionary master graph in Figure 2:

- from microscopic protists >470 Mya ($\Phi_m \approx 10^3$ erg/s/g)
- to gymnosperms ~350 Mya (~$5 \times 10^3$)
- to angiosperms ~125 Mya (~$7 \times 10^3$)
- to highly efficient $C_4$ plants ~30 Mya (~$2 \times 10^4$).

*4.5. Animals*

As widely recognized, both plants and animals engage energy as a vital feature of life. In the animal domain, the principal biological process is respiration, whereby most animals aerobically respire to fuel minimal maintenance (basal metabolic rate) as well as to enhance more active lifestyles (field metabolic rate) when added $O_2$ consumption rises to meet increased demand for ATP production during stress, growth, and thermoregulation (and beyond that, though rarely, catabolic anaerobic pathways, such as glycolytic production of lactic acid, which can generate additional ATP during brief bursts of maximum activity like that experienced by darting lizards and marathon runners).

Sec. 4.4 above examined plants in some detail, suggesting how energy rate density can be reasonably judged as both a complexity metric and evolutionary facilitator. Energy flows in animals are hereby analyzed in a consistent way by imploring the same working hypothesis of $\Phi_m$ as a rational complexity gauge for all ordered systems observed in the Universe—namely, by estimating specific (*i.e.*, mass-normalized) metabolic rates for whole bodies (this section) and networked brains (next section) among a large sample of animals. Some recent research (*e.g.*, [120-122]) has embraced the idea of energy flow as an organizing process in Nature, but these studies are mostly theoretical and restricted to life, thus forsaking empirical metrics (such as $\Phi_m$) of extraordinarily wide scope.

*4.5.1. Evolution and Complexity.* To connect the discussion of plants as (mostly) $O_2$-producers with that of animals as $O_2$-consumers, note that recent studies of Mo isotopes in ocean sediments imply that increased $O_2$ levels beginning around the Cambrian period might have fostered increased size among animals [123]. Two growth spurts in animal biovolume are evident, one ~2.3 Gya (termed above the Great Oxygenation Event) when $O_2$ first began accumulating on Earth (yet was then only a few percent of its abundance today), and the second ~600 Mya when organisms emerged from their microscopic world and began developing skeletons and shells (yet $O_2$ still totaled only ~10% of atmospheric gases). A third stepwise oxygenation of the ocean (and by implication a further doubling of $O_2$ gas) probably occurred in the Devonian ~400 Mya, when the development of plant ecosystems roughly correlate with the increasing size of fossil predatory fish. Climate change, variable glaciation, and ozone buildup also likely contributed to animal growth, but the message seems clear: Larger animals use more energy, that is require more $O_2$, and such a requirement can probably be met only in $O_2$-rich waters.

A wealth of paleontological and genetic data available today imply that animals (multicellular eukaryotes) *generally* became increasingly complex with time—both in structure and function of individual organisms as well as in organization of ecological communities—indeed dramatically so in the Phanerozoic Eon since the Cambrian Period [124-127]. A clear yet rambling succession of life-forms, broadly identifiable yet minus transitional details, is evident during the past ~0.54 Gy: invertebrates (>500 Mya), fish (~500 Mya), amphibians (~365 Mya), reptiles (~320 Mya), mammals (~200 Mya), and birds (~125 Mya). Much as suggested for plant evolution in Sec. 4.4, energy flow potentially affected animals, linking complexity growth and evolutionary pathways with increased energy usage, all of it broadly in accord with the Darwinian precept of descent with modification guided by biological selection—from ectotherms in the hot, damp climates of the Palaeozoic Era, to increasingly diverse animals of intermediate metabolism that thrived in the warm and drier Mesozoic, and then to endotherms in the cooler, fluctuating climates of the Cenozoic.

Much of this change occurred by means of random evolutionary opportunities to secure food and escape predation, which initially required transport of $O_2$ reserves from the open waters and thus metabolically elevated levels of energy consumption, followed by the terrestrialization of the vertebrates that required yet more energy largely because reptiles moved on legs and pumped their chests. Mammalian adaptation further aided the rising complexification of the animal world, resulting in not least the emergence of energy-hungry primates, including our high-energy human society, the last of these discussed in Sec. 4.7 below and especially in [22]. While there is no evidence that any of these energy additives were goal-directed, each arguably presented adaptive advantages for some species throughout a long and meandering evolutionary process during the most recent 10% of Earth's history.

*4.5.2. Ectotherms.* Ectothermic (also known as poikilothermic) animals control their body temperature (~22°C) by means of external heat sources and include both invertebrates (all arthropods, including insects, worms, crustacea, and their relatives) as well as lower vertebrates (fish, amphibians, and reptiles). As a group, ectotherms have less active metabolisms compared to endotherms that include mammals and birds and that self-regulate their core body (37-42°C generally, which is higher than the normal 37°C [98.6°F] for most mammals, possibly to ward off fungi) by digesting food [128]. In fact, low metabolic rates are notably characteristic of all extant reptilian taxa, which were the first fully terrestrial vertebrates and which later gave rise (probably along independent lines of descent during the early Mesozoic) to two major phylogenetic radiations of endothermic mammals and birds. Cold-blooded ectotherms also have lower specific metabolic rates, hence lower values of $\Phi_m$, than their warm-blooded cousins. Here, *in vitro* $O_2$ consumption rate effectively estimates metabolic rate, but caution is advised regarding wet and dry body mass, for it is wet (living) mass that counts when deriving values of $\Phi_m$ in a consistent manner among all living creatures. Furthermore, it is the basal rate (for fasting, resting, inactive states) that is most telling when comparing $\Phi_m$, and not the more active rates experienced when contending with all the challenges of relying on the environment (as do ectotherms) or finding enough food (endotherms) to maintain body temperature [129]. Added care is also required regarding incompatible units found throughout the bioscience literature; although the thermodynamic (cgs-metric) units used here may be unfamiliar to some researchers, these same units are applied consistently and uniformly throughout this review of physical, biological, and cultural systems: thus 1 liter of $O_2$ consumption equals $\sim 2\times 10^{11}$ erg or ~4.8 kcal [130].

Current metabolic data are insufficient to show any clear evolutionary differences in $\Phi_m$ values among the ectotherms [131]. Variations are statistically indistinguishable among the lower vertebrates, including fish, amphibians, and reptiles; most of their $\Phi_m$ values range between $2\times 10^3$ and $10^4$ erg/s/g, with a mean of $\sim 4\times 10^3$ erg/s/g. As expected from paleontology, aerobic capacities were not appreciably expanded as animals made the transition to land; reptiles and amphibians have no more energy needs than fish of comparable size. Among invertebrates, which are also ectothermic and constitute >95% of all animal species, $\Phi_m \approx 10^4$ erg/s/g $\pm$ ~30%; their slightly higher $\Phi_m$ than for the lower vertebrates, if significant, may owe to some invertebrates being active flyers, including minute insects, which likely require more power per unit mass (as do birds, see below). That these mean values are only slightly higher than for some photosynthesizing plants (*cf.*, Sec. 4.4) is not surprising. The resting rates for the least evolved respiring ectothermic animals are not likely much more complex than efficiently photosynthesizing land plants, these two biological advancements having matured roughly contemporaneously during the Paleozoic. Occasional outliers and minor overlaps in $\Phi_m$ values are evident throughout the evolutionary record for comparably complex life-forms, as acknowledged here and discussed in Sec. 5.8.

*4.5.3. Endotherms.* In contrast to the ectotherms, warm-blooded endotherms (also known as homeotherms) have distinctly higher levels of specific metabolism, hence higher values of $\Phi_m$. Many field studies and laboratory measurements of animals having comparable body mass and temperature show basal metabolic rates 5-20 times greater in mammals than in reptiles [131-133]. Three-quarters of all known mammals display a range in $\Phi_m = 10^4 - 10^5$ erg/s/g, with a mean of $\sim 4\times 10^4$ erg/s/g. Variations in metabolic rates among mammals are apparent throughout these data; besides the most dominant influence of differing body mass, such variations likely reflect environmental conditions, ongoing adaptation, and numerous other ecological factors that influence metabolism such as habitat, climate, diet, and taxonomy [134-136]. To give a few examples: seals and whales have $\Phi_m$ values about twice those of other animals of their size because they need to thermoregulate their bodies in cold water; small desert mammals have lower $\Phi_m$ values than others of their size because they have adapted to a scarcity of food and water; and placental mammals have typically thrice the $\Phi_m$ value of similarly sized marsupials because they are viviparous and have extra layers of energetically expensive brain mass.

Caution is needed to distinguish between basal (minimum) and active (vigorous) metabolic rates [137, 138] since the two can be as different as the fuel consumption of an automobile idling at a traffic light or speeding along a highway. For example, a horse expends ~$5 \times 10^5$ erg/s/g at maximum exertion, ~$3 \times 10^5$ during regular exercise, yet only ~$8 \times 10^3$ at rest [139]; cheetahs (the fastest land animal) achieve even higher $\Phi_m$ values (~$10^6$ erg/s/g) while briefly accelerating during hunting [140]; even slower yet ravenous black bears can exceed $10^5$ erg/s/g when fattening up each fall by foraging for berries ~20 hours daily, but then hibernate for months with $\Phi_m$ values orders of magnitude lower. Overall, laboratory studies of sustained (field) metabolic rates typical of all free-living animals in the wild display enhancements in $\Phi_m$ by factors of 3-10 (and up to 50 for maximum exertion) over their basal rates, yet still reveal that mammals outpace reptiles by nearly an order of magnitude [138, 141]. The different rates can nearly overlap for disparate life-forms, much as noted two paragraphs above for simple animals (heterotrophic ectotherms) and efficient plants (advanced photoautotrophs). Likewise, endothermic vertebrates at rest and ectothermic insects in flight display comparable metabolic levels, as do maximum $\Phi_m$ for darting reptiles when compared to many resting mammals. However, mixing metabolic rate states creates unfair comparisons and bewildering confusion in the literature does not help. When level assessments are made for the same type of specific metabolic rate, relative $\Phi_m$ values are clear and unambiguous: higher vertebrates (mammals and birds) have greater energy rate densities than any of the lower vertebrates or invertebrates.

*4.5.4. Birds.* Also endothermic, birds evolved from carnivorous, feathered dinosaurs during the late Mesozoic (~125 Mya), and among vertebrates have the highest values of $\Phi_m \approx 10^5$ erg/s/g, which can sometimes reach nearly an order of magnitude greater during sustained flight or while earnestly foraging for food for their nestlings. Such high $\Phi_m$ implies that birds' normal metabolisms are more energetically comparable to active (not basal) metabolisms among non-fliers; estimates of basal rates for birds resting at night, which would provide legitimate comparisons, are scarce and anecdotal. Many passerine (perching, frugivore) birds have $\Phi_m \approx 5 \times 10^5$ erg/s/g, which is ~30% higher ([142] claims this, but [138] refutes it) than non-passerine fliers whose energy rate densities are comparable to mammals when active; however, uncertainties linger about reported avian rates being basal, active, or some sort of operational average. Hummingbirds, for example, when actively hovering can use as much as 8 times more energy than their resting rate, yet while sleeping (more than half of each day) their rates decrease to ~3 times less than basal when their body temperature drops to nearly that of the surrounding air; the former state requires them to ingest nectar daily equal to ~50% of their body mass, while the latter subsides on minimal energy stores. Murres, which are penguinlike seabirds, expend more energy per time in flight than any other bird ($\Phi_m \approx 10^6$ erg/s/g); this active rate, however, exceeds by a factor of ~30 times their much lower basal rate at rest (~$3 \times 10^4$ erg/s/g), which is more representative of average avian metabolic rates (since they rarely fly) and is probably why penguins long ago opted for swimming than flying as the latter is too expensive [143]. Basal-active comparisons can also be made for mammals, such as for humans who maintain our basal rate by ingesting food daily equal to ~3% of our body mass; yet our active metabolisms also increase by more than an order of magnitude above our basal rates when swimming, jumping, or running (see section on humans below), for which $\Phi_m$ averages $2 \times 10^5$ erg/s/g [144]. For nearly all active fliers <1 kg, $\Phi_m$ is less than for comparably massive mammals while running; generally, active land mammals have similar $\Phi_m$ values to those of most airborne species. Furthermore, birds, much like human marathoners and cyclists who consume many times their normal food intake (up to ~$5 \times 10^5$ erg/s/g, or 3000 W per capita compared to the nominal 130 W for humans), are fueled partly by rapid expression of bodily energy reserves (anaerobic glycolysis), not by sustained, concurrent energy intake; these enhanced metabolic rates are atypical physiologically, hence their more representative rates are lower when averaged over time.

In addition to their habitually active states, birds might also have high values of $\Phi_m$ partly because they are conceivably more complex than most other animals, including humans. After all, birds normally operate in three-dimensional aerial environments, unlike much of the rest of animalia at the two-dimensional ground level; thus avian *functions*, quite apart from structural integrity, might be legitimately considered, somewhat and sometimes, more complex than those of the rest of us who cannot fly [17]. Brains aside (*cf.*, Sec. 4.6), the bodies of fliers can arguably be judged more complex than non-fliers, given the former's intricate lung sacs, pectoral muscles, and wing aerofoils that allow a constant, one-way flow of $O_2$-rich air that helps birds maintain high metabolic rates to generate enough energy for flight. The European swift bird, for example, can fly non-stop up to $10^3$ km during breeding season, performing many functions including sleeping on the way; foraging bumblebees can fly several km/day from their hives, traveling up to 10 m/s (~30 km/hr) while flapping their wings 160 times *per second* and not surprisingly sporting large appetites during powered flight. The act of flying does indeed demand great skill, more energy (to work against gravity), and a higher cost of living in general, requiring birds to master (effectively) spatial geometry, aeronautical engineering, molecular biochemistry, and social stratification. Avian species are impressive by any measure; their speed, maneuverability, and endurance are outstanding among all known life-forms, so perhaps they *should* have large values of $\Phi_m$. Furthermore, $\Phi_m$ computations suggest an even wider, more interesting, trend: not only birds among vertebrates, but also insects among invertebrates (see above) and aircraft among machines (see Sec. 4.7) all have the highest energy rate density within their respective categories, almost certainly because they operate in three dimensions. That doesn't make fliers smarter than us, merely their functions are arguably more complex (when flying) than nearly anything humans biologically do in two dimensions.

Thinking indeed broadly, extraordinary avian physiology might resemble not only high-endurance athletes, but also enhancements in galactic ecology. Each category of system—animals and galaxies—includes minority members with exceptionally high metabolisms during short periods of maximum exertion when power expenditures climb substantially. Both in-flight birds and on-race marathoners, which while temporarily sporting their most active states have among the highest animal complexity levels, resemble the extreme energetics of briefly erupting active galaxies (*cf.*, Sec. 4.1); each tops the charts of specific metabolic rates within their respective classes, as values of $\Phi_m$ climb several factors higher than their basal, or normal rates.

*4.5.5. Complexity Rising.* The central challenge for zoology is to explain the extraordinary diversity of animal species on Earth. In general, major evolutionary stages of life are evident: in turn, protists, plants, reptiles, mammals. Yet can we become more quantitative, numerically analyzing animals in ways similar to that done earlier for galaxies, stars, planets, and plants? The answer seems to be affirmative, but for now and paralleling the brief description of plant evolution in Sec. 4.4, here is a condensed, qualitative outline of the main zoological changes in post-Cambrian times that display increased energy-expenditure levels (adapted from [145]):

The mid-Cambrian (~520 Mya) was characterized by burrowing worms (especially the segmented marine coeloms, compared to their soft-bodied flatworm precursors moving only on the sediment surface) that developed hydrostatic skeletons and associated muscles to exert mechanical leverage—much of it probably an evolutionary advantage to escape from predators, yet which required transport of their own $O_2$ reserve from the open waters and thus an elevated consumption of metabolic energy. By the end of the Silurian (~420 Mya) and well into the Devonian (~380 Mya), several classes of fish-like vertebrates are found fossilized in brackish estuaries and fresh water deposits. For such organisms to adapt to changing salinity and chemical compositions, they likely maintained a stable internal osmotic medium, and the energy cost of such osmoregulation is high; only the mollusks, annelids, arthropods, and vertebrates invaded the nutrient-rich estuaries, which in turn acted as evolutionary corridors leading to colonization of the continents. The result was the rise of reptiles and amphibians, radiating wildly in global diversity as many new, fragmented habitats emerged after vast tracts of tropical forests died, ~305 Mya, probably owing to climate change that dried up those rainforests.

Throughout the Mesozoic (~250-65 Mya), the adaptation of the arthropods, predominantly the insects with their solar-aided metabolic activity, was very successful, yet all insects, which followed the plants onto the land, remained small as predator vertebrates in turn tracked them landward. Further in turn, it was the feeding on ants and termites (myrmecophagy) that supplied the needs of primitive, insectivorous mammals, indeed which still provides the large energy needs of modern shrew-like animals that feed constantly in order to maintain their endothermy. Of special import, the oldest mammals—mouse-sized and insect-eating—evolved from reptiles (therapsids) ~200 Mya.

The terrestrialization of the vertebrates was more complicated, but it too required more energy. Briefly and especially during the globally warm, 80-My-long Cretaceous (the longest geological period surrounding ~100 Mya), the ectothermic herbivores (including the dinosaurs) needed more energy if only because they were moving on legs and bloating their lungs. Early endothermic mammals, greatly restricted during the Cretaceous, flourished as the world entered the Tertiary beginning ~65 Mya, and although initially far from modern mammals, energy requirements rose again. The high and constant body temperature as a mammalian adaptation to terrestrial environments also allowed sophisticated neural processing and complex learned behavior—two of the most prominent breakthroughs resulting from the thermodynamic evolution of the animal world—culminating (at least for now) in the rise of the great apes in the Miocene (~20 Mya), thence on to present high-energy-cost humans with their even higher-energy-utilizing brains (*cf.*, Sec. 4.6). As noted above and stressed again, there is no evidence that these energy enhancements were goal-directed, rather each seems to have granted selective advantages for many sundry species at each and every step of the twisting and turning evolutionary process.

*4.5.6. Allometric Scaling.* Quantitative reasoning in this section on zoology is independent of the ongoing debate about allometric scaling of metabolism among mammals from mice to elephants (spanning 6 orders of magnitude in body mass). Nor is it important here whether their mass-dependent, $M^n$, metabolic exponent $n = {}^2/_3$ as expected from surface-to-volume scaling for spherical bodies dissipating heat from their surfaces [146], or ${}^3/_4$ based on laboratory measures [147] and fractal theory of nutrient supply networks for elastic machines having muscular systems and skeletal loads subject to gravity [148, 149]; in fact, it might be neither [150], as metabolic-rate dependence on body mass likely differs with real-time activity level [151], during lifetime development [152], and among evolutionary lineage [153]. In any case, claims of a universal law of bioenergetics for all life-forms from bacteria to elephants [154] are mathematically [155] and empirically [131] dubious.

Despite these ongoing biological controversies that are further troubled by many exceptions to any such proposed biological "law," all animals, and not just mammals but including a wide range of known heterotrophic species, have specific metabolic rates within a relatively narrow range of $\Phi_m$ extending over a factor of only ~30. The great majority of specific metabolic rates for animals vary between $3 \times 10^3$ and $10^5$ erg/s/g, despite their masses ranging over ~11 orders of magnitude from fairy flies to blue whales [131]; all of their $\Phi_m$ values fall midway between smaller botanical values for photosynthesizing plants (see Sec. 4.4) and higher neurological ones for pensive brains (see below). Among mammals alone, specific metabolic rates vary inversely yet weakly with body mass, $\sim M^{-0.2}$. Throughout cosmic evolution, it is the *specific* metabolic rate that matters most, namely those energy flows that are normalized to mass and that for many life-forms vary weakly with mass, likely as $M^{3/4}/M = M^{-1/4}$. This quarter-power scaling tendency is pervasive in biology, probably the result of physical constraints on the circulatory system that distributes resources and removes wastes in bodies, whether it is the geometrical pattern of blood vessels branching through animals or the vascular network nourishing plants. Biological selection has apparently optimized fitness by maximizing surface areas that exchange nutrients while minimizing transport distances and times of those nutrients [156]. That the smallest animals have somewhat higher $\Phi_m$ values probably owes to their frequent eating habits, high pulse rates, robust activity levels, and relatively short life spans; they live fast and die young. By contrast, the largest animals have slightly lower $\Phi_m$ owing to their more specialized cells, each of which has only limited tasks to perform and energy needed, thus granting greater efficiency and a longer life.

Naturally, those species whose individuals enjoy greater longevity are also likely to experience more extreme environmental stress and therefore be exposed during their longer lifetimes to enhanced opportunities for adjustment and adaptation—and thus for evolution toward greater complexity (as well as devolution toward simplicity and even extinction should those stresses be great). The result with the passage of time, as a general statement for bodies of similar mass, is a feedback process whereby those successful systems able to assimilate greater energy flow live longer, evolve faster, and generally complexify, which, often in turn, leads to higher metabolic rates, and so on.

Deep into discussion of biological metabolism, we once again encounter a widespread astronomical factor—the mass-based gravitational force so integral to our earlier analysis of the underlying agents that spawned increased complexity of, for example, stars and galaxies. Astrophysics and biochemistry are not uncoupled parts of the cosmic-evolutionary scenario, as allometric scaling suggests.

*4.5.7. Humans.* Our bodily selves deserve more than a passing note in any study of complex systems, not because humans are special but because we are them. Each individual adult, globally averaged today (although rising obesely), normally consumes ~2800 kcal/day (or ~130 W) in the form of food to fuel our metabolism. This energy, gained directly from that stored in other (plant and animal) organisms and only indirectly from the Sun, is sufficient to maintain our body structure and warmth (37°C) as well as to power our physiological functions and movements during our daily tasks. (Note that the definition of a thermodynamic calorie, 1 cal = $4.2 \times 10^7$ erg—the amount of heat needed to raise 1 g of H$_2$O by 1°C—does not equal a dietician's large Calorie with a capital "C," which is 10$^3$ times more energetic than a physicist's calorie.)

Metabolism is a dissipative process—a genuinely thermodynamic mechanism. Heat is generated continuously owing to work done by the tissues among the internal organs of our bodies, including contracting muscles that run the heart, diaphragm, and limbs, ion pumps that maintain the electrical properties of nerves, and biochemical reactions that dismantle food and synthesize new tissue. The flow of energy in our human bodies is apportioned among movement (15W), labor (20W), and metabolism (95W), the last of these further subdivided into the brain (20W), gastrointestinal track (20), heart (15), kidneys (8), muscles (15) and other organs (17)—all of which totals 130 W.

Therefore, with an average body mass of 65 kg, a generic adult (male or female) maintains $\Phi_m \approx$ 2x10$^4$ erg/s/g while in good health. Those who consume more, such as residents of the affluent United States (where the daily per capita consumption grew from ~3100 kcal in 1970 to ~3500 kcal in 1995 [157]), usually have larger bodies, thus their $\Phi_m$ values remain ~2x10$^4$ erg/s/g, much as for smaller adults who often eat less. Humans have mid-range mammalian metabolic values because our bodies house average complexity among endothermic mammals, all of which comprise comparable intricacy; all mammals, and not just us, have hearts, livers, kidneys, lungs, brains, muscles, and guts. Despite our manifest egos, human beings do not have the highest energy rate density among animals, nor are our bodies demonstrably more complex than those of many other mammalian species.

The energy budget derived here for humans assumes today's typical, sedentary citizen, who consumes ~65% more than the basal metabolic rate of 1680 kcal/day (or $\Phi_m \approx$ 1.2x10$^4$ erg/s/g) for an adult fasting while lying motionless all day and night. By contrast, our metabolic rates increase substantially when performing occupational tasks or recreational events—again, that's function, not structure. And once again, $\Phi_m$ scales with the degree of complexity of the task or activity. For example, fishing leisurely, cutting a tree, and riding a bicycle require about 3x10$^4$, 8x10$^4$, and 2x10$^5$ erg/s/g, respectively [144]. Clearly, sawing and splitting wood or balancing a moving bicycle are complicated functions, and therefore more energetically demanding activities, than waiting patiently for fish to bite. Thus, in the biological realm, the value-added quality of functionality does indeed count, in fact quantitatively so. Complex tasks actively performed by humans on a daily basis are typified by values of $\Phi_m$ that are often higher than those of even the metabolically imposing birds, in part because birds cannot operate machines or ride bicycles!

Human specific metabolic rates then lie near, but not atop, the upper part of the master curve of rising complexity (Figure 2)—within a lower bound (basal rate) that is midway for most mammals and an upper bound (active rate) typical of most birds in flight. Later, we shall encounter even higher energy rate densities for humanity collectively, that is, for integrated society and its invented machines, both of which are advancing culturally (*cf.*, Sec. 4.7).

Sanity checking yet again, this is how humankind, like all members of the animal world, contribute to the rise of entropy in the Universe: We consume high-quality energy in the form of ordered foodstuffs, and then radiate away as body heat (largely by circulating blood near the surface of the skin, by exhaling warm, humidified air, and by evaporating sweat) an equivalent amount of energy as low-quality, disorganized infrared photons. Like the stars and galaxies, we are indeed dissipative structures as are all Earthly life-forms, thereby making a connection with previous thermodynamic arguments that some researchers might (wrongly) think pertinent only to inanimate systems.

*4.5.8. Energy Rate Densities for Animals.* Consider some representative animals for which metabolic rates are known, noting that those rates vary upwards under stress and exertion; their total energy budgets depend largely on energetically expensive internal organs such as kidneys, hearts, brains, and livers. Laboratory measurements of sustained metabolic rates for 50 vertebrate species [141] found that reptiles, mammals (including rodents, marsupials, and humans), and birds average $\Phi_m \approx$ 9,000, 56,000, and 78,000 erg/s/g, respectively. These and other measures quoted above imply that specific metabolic rates of cold-blooded ectotherms are only a fraction of those of similarly massive warm-blooded endotherms, much as expected on evolutionary grounds. This is hardly surprising since, for endotherms to carry with them portable, thermally regulated bodily habitats, an inevitable energy cost results; the ability to thermoregulate likely confers a competitive, even survival-related, evolutionary advantage, and energy is needed to make it work.

The order-of-magnitude difference in specific metabolic rates among birds, mammals and comparably sized reptiles can legitimately be cast in terms of relative complexity, since the need for endotherms to homeostatically control body temperature (both heating and cooling) is surely a more complicated task that ectotherms simply cannot manage to do—and it is extra energy that allows for this added feature, or selective advantage, employed by birds and mammals over the past few hundred million years. Even so, ectotherms are much more abundant, both in species numbers and in total life-forms, implying that they, too, are quite successful in their own more limited realms.

Among the eukarya (life's 3$^{rd}$ domain that includes all plants and animals), ectotherms have $\Phi_m$ values between $2 \times 10^3$ and $10^4$ erg/s/g, whereas endotherms have not only a similarly wide range of values but also higher absolute values, namely $10^4 - 10^5$ erg/s/g. The former are clearly among the earliest of biological evolution's animal creations, whereas the latter are widely considered more advanced, indeed among the most complex, of Nature's many varied life-forms; with their mobile microenvironments (shelter, fire, clothing, *etc*), the endotherms have enjoyed a strong competitive advantage, enabling them to adaptively radiate into even the most inhospitable parts of Earth's biosphere.

Figure 7 summarizes values of $\Phi_m$ for the whole bodies of a spectrum of mature, adult animals within the biological-evolutionary phase of cosmic evolution. These are mean values for a wide range of diverse taxonomic groups that are resting (basal) with normal body temperature, culled, computed, and averaged from many references noted above. Evolutionary times approximate those at which various animal types emerged in natural history, albeit only since the Cambrian ~540 Mya. This entire graph fits within the mid-to-upper part of the master curve of rising complexity plotted in Figure 2.

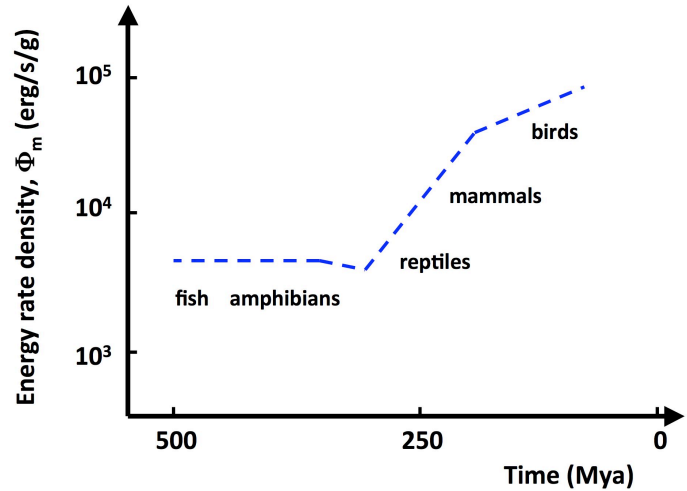

FIGURE 7: The complexity of vertebrates, expressed in terms of $\Phi_m$, is shown here rising in order to highlight some of the increasingly intricate structures and functions for a variety of animals at various stages of biological evolution. Note how endotherms (including mammals and birds) have higher energy rate densities than ectotherms (including invertebrates as well as lower vertebrates such as fish, amphibians and reptiles) among all taxonomic groups found on Earth.

*4.5.9. Evolutionary Advancement.* No strong correlations between $\Phi_m$ values and biological evolution are evident for individual members of the animal kingdom. Energy rate density may well qualify as a broad complexity metric for life, but current data preclude $\Phi_m$-related statements about specific evolutionary paths for discrete species within major taxonomic groups. Suffice it to say that nearly all zoological $\Phi_m$ values are tightly confined to within hardly more than an order of magnitude of one another, nestled midway between smaller botanical values for photosynthesizing plants (see Sec. 4.4) and higher neurological ones for central nervous systems (see Sec. 4.6). Nonetheless, correlations do link evolution, complexity, and $\Phi_m$ for major animal categories, notably those separating reptiles, mammals, and birds. For example, endothermy is surely one of the most striking animal adaptations, requiring extensive restructuring of many parts (including lung, heart, and skeletal muscle) of vertebrate bodies. The greater aerobic heat production in the endotherms is the basis of their homeothermic condition that grants them independence from environmental thermal fluctuations, and this arguably makes them more complex. Endothermy likely evolved in mammals from reptiles in the early Mesozoic as mitochondrial volume density gradually increased in their respective tissues, causing microscopic metabolisms to accumulate and with them total organismal specific metabolic rates to rise [158]. The original vertebrates (possibly ostracoderms) were active, predatory carnivores with metabolic signatures similar to most modern fish; the transition of vertebrates from aquatic to terrestrial habitats eventually enabled greater $O_2$ use, since $O_2$ in the aerial environment is more easily accessible given its increased diffusivity and concentration. However, most traits related to $O_2$ consumption do not fossilize and other factors have also been implicated as having granted major selective advantages [159]. Thermoregulation itself allows body temperature of mammals and birds to remain both higher and more constant than those of most ectothermic vertebrates, and this alone might enhance prospects for survival; endothermy, with its portable microenvironment, surely conferred competitive evolutionary advantages in benign environments and allowed those species so endowed to adaptively radiate into hostile parts of the biosphere. Higher levels of $O_2$ consumption also likely expanded the range of sustainable exertion and long-distance endurance, granting opportunities for greater complexity to parallel the rise in $\Phi_m$ values for mammalian and avian lineages. Regardless of how it emerged, a clear prerequisite underlies endothermy: more energy is required to attain it.

This is not to assert that energy, solely and exclusively, drives biological evolution. Energy flow is probably only partly responsible for evolutionary advancement of rising complexity. Non-evolutionary effects also surely contribute to the observed range in $\Phi_m$ values since stressful environments can push some organisms to extremes. For example, aquatic mammals have specific metabolic rates that are necessarily higher (by factors of 2-3) than those of similarly sized land mammals (since, much as for birds, the former operate in three dimensions, in this case where water conducts heat 20 times faster than air). Opposite extremes are found in desert mammals, whose anomalously low specific metabolic rates reflect food shortages, though they can rehydrate rapidly by drinking the equivalent of a third of their body weight in minutes. Dietary, hydration, behavioral, and habitat factors all likely cause variations in $\Phi_m$ values in addition to evolution per se, resulting in rare outliers in such diverse samples of animals. Body mass itself seems the biggest cause of variation among metabolic rates for mammals; much the same is true for birds, as body mass alone accounts for >90% of their variation in $\Phi_m$ [142]. All things considered, macroscopic life-forms display clear and abiding, yet general, trends between evolution-associated complexity and energy rate density.

*4.5.10. Animal Summary.* Animals among biologically complex systems regularly exhibit intermediate values of $\Phi_m = 10^{3.5-5}$ erg/s/g—and human bodies rightfully are not the most complex among them. So much for human uniqueness; all animals are outstanding in their own ways, and although we do have special traits, so do salmon, giraffes, robins, and other large vertebrates. Onward across the bush of life (or the arrow of time); much the same temporal trend of rising $\Phi_m$ holds for adult, respiring animals while evolving and complexifying:

- from fish and amphibians 370-500 Mya ($\Phi_m \approx 4 \times 10^3$ erg/s/g)
- to cold-blooded reptiles ~320 Mya (~$3 \times 10^3$)
- to warm-blooded mammals ~200 Mya (~$4 \times 10^4$)
- to birds in flight ~125 Mya (~$9 \times 10^4$).

To sum up the past two sections on plants and animals, the rise of $\Phi_m$ *generally* parallels the emergence of major evolutionary stages on the scale of life's history: eukaryotic cells are more complex than prokaryotic ones, plants more complex than protists, animals more complex than plants, mammals more complex than reptiles, and so on. Claims regarding the role of $\Phi_m$ in evolutionary advances are broad and general, not specific and detailed along individual lineages; the objective in this research program is to identify how well life-forms fit quantitatively within the larger scenario of cosmic evolution. Similarities between galaxies and animals (as briefly noted earlier) are amply evident, including variation within category types, adaptation (or adjustment) to changing conditions, and possibly natural (*i.e.*, physical not Darwinian—*cf.*, Sec 5.6) selection among interacting galaxies [160], much as proffered above for stars and plants as well. All these systems are open to their environments, with matter and energy flowing in while products and wastes flow out, indeed all resemble metabolisms at work on many scales. Whether stars, galaxies, or life itself, the salient point seems much the same: The basic differences, both within and among Nature's many varied systems, are of degree, not of kind. We have discerned a common basis upon which to compare hierarchically all material structures, from the early Universe to contemporary Earth—again, from big bang to humankind inclusively.

*4.6. Brains*

Regarding brains, which nuclear magnetic resonance (fMRI) imaging shows are always electrically active regardless of the behavioral posture of their parent animal bodies (even while completely resting), they too derive nearly all their energy from the aerobic oxidation of glucose in blood; thus, for brains, basal and active rates are comparable (with blood flow in an idle brain ≤10% lower than during task-based activities). General trends in rising complexity noted above for bodies are also evident for brains, although with higher $\Phi_m$ brain values for each and every animal type—much as expected since cerebral structure and function are widely regarded among the most complex attributes of life [161, 162]. Here, quantitative details are compiled from many sources, again treating brains as open, non-equilibrated, thermodynamic systems, and once more casting the analysis of energy flow through them in terms of energy rate density. (While acknowledging several other potentially useful neural metrics—cortical neuron numbers, encephalization quotients, brain/body ratios [163]—I specifically examine brains here for their $\Phi_m$ values in order to be scrupulously consistent with my proposed complexity metric of energy rate density for all complex systems.) However, brain metabolic values gathered from the literature often suffer, as noted above for bodies, from a lack of standard laboratory methods and operational units; many reported brain masses must be corrected for wet (live) values (by multiplying measured *in vitro* dry masses by a factor of 5 since *in vivo* life-forms, minus their bones and including brains, are ~80% $H_2O$). Note also that the ratio of brain mass to body mass (used by some neuroscientists as a sign of intelligence) differs from the ratio of brain power to brain mass (which equals $\Phi_m$), nor is the term "brain power" the same as that often used in colloquial conversation, rather here it literally equals the rate of energy flowing through the cranium.

*4.6.1 Energy Rate Density for Brains.* No attempt is made to survey brains comprehensively, rather only to analyze their energy budgets broadly; representative mean values of brain $\Phi_m$ suffices for a spectrum of extant animals. Comparing mammals and reptiles, $\Phi_m \approx 10^5$ erg/s/g for mice brains (in contrast to ~$4 \times 10^4$ for their whole bodies) exceeds ~$5 \times 10^4$ erg/s/g for lizard brains (~$3 \times 10^3$ for their bodies [133]); this is generally the case for all such animal taxa as $\Phi_m$ values are somewhat greater for mammal brains than those for reptile brains by factors of 2-4, and those for mammal bodies by roughly an order of magnitude [164]. The great majority of vertebrate fish and amphibians show much the same 5-10 times increase in brain over body $\Phi_m$ values [165, 166] with, as often the case in biology, some outliers [167]. Even many invertebrate insects show several factors increase in $\Phi_m$ values for their brains (~$5 \times 10^4$) compared to their bodies (~$10^4$), most notably the flying insects [168]. However, for brains in particular, ectotherms generally have only slightly lower values of $\Phi_m$ than endotherms, the reason being that on a cellular level brains function in essentially the same way for both warm- and cold-blooded creatures and heat production plays a relatively minor role in brain energy expenditure [169].

Among mammals alone, primates, which evolved from tree-dwelling, insect-eating ancestors ~65 Mya, have not only high brain/body mass ratios but also relatively high $\Phi_m$ values (~$2\times10^5$ erg/s/g) for those brains. Although primates allocate for their brains a larger portion (8-12%) of their total bodily (resting) energy budget than do non-primate vertebrates (2-8%) [164, 170-171], average primate brains' $\Phi_m$ values tend to be comparable to those of brains of non-primates; brain mass-specific, allometric scaling is even slighter— $M^{-0.15}$ —than for bodies of animals as noted in Sec. 4.5, causing $\Phi_m$ brain values to remain approximately constant across 3 orders of magnitude in mammalian brain size [172]. As with bodies above, brains do not necessarily confer much human uniqueness; brains are amazing, but all animals have them, and our neural qualities seem hardly more than linearly scaled-up versions of those of other primates [173].

Brains of birds are also revealing, although the derisive term "birdbrain" is quite unfair to some avian species that demonstrate remarkable cognition [174]. Brains of birds average an order of magnitude larger than those of equivalently massive reptiles. Brain/body mass ratios for the cleverest birds, such as crows and ravens that display much intraspecies cooperation and social cunning, are comparable to those of some primates. Brain $\Phi_m$ values are also comparable, again because less energy of a bird's total body metabolism is devoted to its brain, probably owing to the formidable energetic requirements of bodily flight. As noted in the next paragraph, the most evolved primates (especially humans) direct to their brains as much as a quarter of their total body metabolisms, whereas birds, like all other animals, allocate much less. Such subtle differences between brain/body ratios and relative $\Phi_m$ brain comparisons might imply that the latter could be a better sign of intelligence—if only data were available.

*4.6.2. Human Brain.* Adult human brains—without any anthropocentrism implied, among the most exquisite clumps of living matter in the known Universe—have cranial capacities of ~1350 g and require ~400 kcal/day (or ~20 W) to function properly. Thus, while thinking, our heads glow in the infrared with as much energy as a small lightbulb; when that "bulb" turns off, we die. Our brains therefore have $\Phi_m \approx 1.5\times10^5$ erg/s/g, most of it apparently to support the unceasing electrical activity of ~$10^{11}$ neurons. ($\Phi_m$ computations reveal that human hearts and digestive tracts are similarly complex, and perhaps rightly so regarding vital structure and functionality needed to survive; if you lose an arm or leg you wouldn't die, but if a brain, heart, or gut is lost you would, thus comparable complexities among some bodily organs are not surprising.) Such brain power per unit mass flowing through our heads is larger than for any living primate—not merely ~10 times higher $\Phi_m$ than for our bodies, but also slightly higher than for the brains of our closest living evolutionary cousins, namely the great apes, including chimpanzees. This substantial energy-density demand testifies to the disproportionate amount of worth Nature has invested in evolved human brains; occupying only ~2% of our total body mass yet accounting for 20-25% of our body's total energy intake (as measured by $O_2$ consumption [175]), our cranium is striking evidence of the superiority, in evolutionary terms, of brain over brawn.

Furthermore, our central nervous system's share of our total (basal) metabolic budget—the just mentioned ≥20% of our daily bodily energy intake—means that we devote 2-10 times greater percentage of our body metabolism to our brain than any other anthropoid. The great apes (anthropoid primates) devote only 7-12% typically, other mammals (vertebrates such as rats, cats, and dogs, but excluding humans and primates) use 2-6%, and reptiles even less. Of particular import, our closely related chimpanzees not only have ~3 times less brain/body mass ratio than do humans, but they also require about half the relative energy allocation of a human brain. In any case and by all accounts, brains everywhere are energy-hungry organisms.

As with all structured systems in the Universe, animate or inanimate, human brains have $\Phi_m$ values that vary somewhat depending on level of development. Although mature adult brains typically consume as much as 25% of a body's total energy consumption, young brains of newborn children can utilize up to 60% of the bodily energy acquired—a not unreasonable finding given that a human's lump of neural mass doubles during the first year of life and synapses grow dramatically in the pre-school years [176]. Thus, $\Phi_m$ averages several times larger for infant brains than for adult brains—a pattern often evident throughout cosmic evolution: Earth, for example, needed substantially more energy to develop its rocky being, but less so now to maintain it; much the same trend pertains to the Sun as its protostellar stage had higher energy rates than its normal fusion today (*cf.*, Sec. 4.2, and 4.3 for stars and planets). Likewise, as implied in Sec. 4.5 for animals, during ontological development many organisms also apparently change from higher to lower metabolic rates.

*4.6.3. Complexity — $\Phi_m$ Correlation.* Complex brains with high $\Phi_m$ values, much as for complex whole animal bodies above, can be *generally* correlated with the evolution of those brains among major taxonomic groups [162]. Further, more evolved brains tend to be larger relative to their parent bodies, which is why brain-to-body-mass ratios also increase with evolution generally—mammals more than reptiles, primates notable among mammals, and humans foremost among the great apes [163, 164].

Relatively big brains are energetically expensive. Neurons use energy as much as 10 times faster than average body tissue to maintain their (structural) neuroanatomy and to support their (functional) consciousness; the amount of brain devoted to network connections increases disproportionately with brain size and so does the clustering and layering of cells within the higher-processing neocortex of recently evolved vertebrates [177, 178]. Much of this accords with the "expensive-brain hypothesis" [179, 180], which posits that high brain/body ratios are indeed more energetically costly (at least for mammals and many birds), that energy flow through brains is central to the maintenance of relatively large brains (especially for primates), and that relatively large brains evolve mainly when they manage to use more energy, often by stealing from other bodily organs or functions.  Although the human brain's metabolic rate is not much greater than for some organs, such as the stressed heart or active kidneys, regional energy flux densities within the brain greatly exceed (often by an order of magnitude) most other organs at rest.

The pressures of social groups and social networking might also direct growth in brain size, cognitive function, and neurophysiological complexity along insect, bird, and primate lineages [181, 182].  Human brain size has increased dramatically during the past few My, in contrast to those of our great ape relatives.  Much fieldwork seeks to understand how the challenges of living in changing environments or even in stable social groups might have beneficially enhanced cognitive abilities among primates, especially humans, during this time.  However, in the spirit of this research program that emphasizes the concept of energetics throughout natural history, the use of energy by brains may well be a contributing factor, and perhaps even a prerequisite, in the evolution of brains and of their increased brain-to-body-mass ratios.  The just-mentioned expensive-brain hypothesis predicts that relatively larger brains evolve only when either brain energy input increases directly from the environment or energy allocation shifts to the brain from another part of the body, such as energy-rich tissues of the digestive tract in primates and the pectoral muscles in birds.

Throughout biology generally, brain tissue is known to be energetically costly, requiring nearly an order of magnitude more energy per unit mass than most other body tissues at rest.  This high-energy toll on the brain might therefore constrain biological selection's effect on an animal's survival and/or reproductive success; in fact, the brain is the first organ to be damaged by any reduction in $O_2$.  Recent data on a large sample of basal metabolic rates and brain sizes among vertebrates do suggest that energy flows through brains are key to the maintenance of relatively large brains, especially for non-human primates.  Furthermore, among ~550 species of mammals, the brain-to-body-mass ratio displays a positive correlation with metabolic rate, and even among ~400 species of birds the expensive-brain hypothesis holds [174].

Among more recent prehistoric societies of special relevance to humankind, the growing encephalization of the genus *Homo* during the past ~2 My might provide further evidence of biological selection acting on those individuals capable of exploiting energy- and protein-rich resources as their habitats expanded [183].  By deriving more calories from existing foods, cooking likely encouraged cultural innovations that allowed humans to support big brains [184].  Heated food does accelerate chewing and digestion, allowing the body to absorb more nutrition per bite; cooking may well be a uniquely human trait.  Energy-based selection would have naturally favored those hominids who could cook, freeing up more time and energy to devote to other things—such as forming social relationships, creating divisions of labor, and fueling even bigger brains, all of which arguably advanced culture. As with many estimates of human intelligence, it is not absolute brain size that apparently counts most; rather, brain size normalized by body mass is more significant, just as the proposed $\Phi_m$ complexity metric is normalized by mass, here for brains as for all complex systems at each and every stage of cosmic evolution along the arrow of time, from primordial Universe to the present.

*4.6.4. Summary for Brains:* Not only are brains voracious energy users and demonstrably complex entities, but evolutionary adaptation also seems to have favored for the brain progressively larger allocations of the body's total energy resources. The observed, *general* trend for active brains *in vivo*, broadly stated though no less true for the vast majority of animals, is that their $\Phi_m$ values are systematically higher than for the bodies that house them. Nearly all brain values fall within a narrow range of $\Phi_m$ values between lower biological systems (such as plants and animal bodies) and higher cultural ones (such as societies and their machines). Although absolute brain masses span ~6 orders of magnitude, or a factor of about a million from insects to whales, their $\Phi_m$ brain values cluster within only a few factors, more or less depending upon their mass and evolutionary provenance, of ~$10^5$ erg/s/g.

*4.7. Civilization*

Energy empowers humans today in countless ways by reducing drudgery, increasing productivity, transforming food, providing illumination and transportation, powering industrial processes, conditioning space for households and buildings, facilitating electronic communications and computer operations, *etc*. To examine how well cultural systems resemble physical and biological systems—and thus to explore cultural evolution within a unifying cosmic context—it is instructive to quantify culture, where possible, by means of the same heretofore concept of energy rate density. I do so largely in order to skirt the vagueness of social studies while embracing once again empirical-based energy flow as a driver of cultural evolution.

*4.7.1. Society Advancing.* Consider modern civilization *en masse*, which can be considered the totality of all humanity comprising an open, ordered, complex society going about its daily business. Today's ~7.3 billion inhabitants utilize ~19 TW to keep our global culture fueled and operating, admittedly unevenly distributed in developed and undeveloped regions across the world (extrapolated from [185]). The cultural ensemble equaling the whole of humankind then averages $\Phi_m \approx 5\times10^5$ erg/s/g, which is about an order of magnitude more than any single human being. As expected, a group of intelligent organisms working collectively is more complex than its individual human components [186]; the influence of group size on cultural complexity is further suggestive from analyses of growing cities [22]. These findings abide by the predictions of the energy-rate-density metric hypothesized earlier [8], and is a good example of the whole being greater than the sum of its parts (*cf.*, Sec. 5.7), a common characteristic of emergence fostered by the flow of energy through organized, and in this case social, systems.

Note that in computing $\Phi_m$ for contemporary society, only the mass of humankind itself is used. The mass of modern civilization's infrastructure—buildings, roadways, vehicles, and so on—is not included, any more than is the mass of the clothes we wear when calculating $\Phi_m$ for the human body, or the mass of bodies themselves when evaluating brains, or the mass of our host Galaxy when evaluating the Sun. That is, human society is taken literally as synonymous with the assemblage of humanity *per se*, since the fundamental building blocks of society are its people; what matters most is the total energy utilized by the human social aggregate. Much the same pertains, for example, when examining an ant colony as a superorganism; such extended systems have dirt, tunnels, and rocks, yet the biological essence of the ordered colony is the total mass of the networked ants. When assessing the degree of system complexity, it is reasonable and proper to analyze ordered systems separately from their disordered environments, which is what has been done consistently and uniformly for all earlier thermodynamic diagnoses throughout this study.

Rising energy expenditure per capita has been a hallmark in the origin, development, and evolution of humankind, an idea dating back decades [187, 188]. However, none of these early energy-centered cultural theses addressed causality or were in any way quantitative, yet some of them did speculate that enhancements of energy within living systems likely result from cultural selection and thermodynamic principles. More recently, analytical use of the $\Phi_m$ diagnostic has been extensively and realistically employed to examine the behavior of the Mayan Indians (including their society's virtual collapse from not only conquest and disease but also inadequate energy management), inferring that life and society (even today) can remain viable provided that evolutionary strategies maintain sustainable energy with a steady flow ultimately from the Sun [189]. In contrast to most cultural studies, the present analysis seeks to specify, even if only broadly, such a causative agent, or prime mover, in the guise of cosmic expansion, which, in turn, orchestrates flows of energy within increasingly evolved, complex systems.

Culture itself is often defined as a quest to control greater energy stores [190]. Cultural evolution occurs, at least in part, when far-from-equilibrium societies dynamically stabilize their organizational posture by responding to changes in energy flows through them. Quantitative assessment of culture, peculiar though it may be from a thermodynamic viewpoint, need be addressed no differently than for any other part of cosmic evolution [191]. Values of $\Phi_m$ can then be estimated by analyzing society's use of energy by our relatively recent hominid ancestors.

*4.7.2. Energy Rate Density for Society.* The following few paragraphs gauge energy usage among a variety of human groups throughout time, illustrating how, in turn, advancing peoples of the genus *Homo* utilized increasing amounts of energy beyond the 2-3000 kcal/day that each person actually eats as food [192-195, 3, 16]. For perspective, first consider members of perhaps the most primitive society of hominids, who had available for work only the physical energy of their individual work ethic. Most published estimates suggest that such ~40-kg australopithecine ancestors ~3 Mya would have consumed ~2000 kcal of food per day, granting each of them $\Phi_m \approx 22,000$ erg/s/g. Quite possibly, >99% of our evolutionary history was spent foraging for food in small bands of a few dozen to a few hundred people.

Hunter-gatherers ~300 kya likely augmented by small amounts the basic energy of food needed to survive. Anthropologists have studied these relatively simple cultures and the energy flowing through them, not only by unearthing ancient habitats of extinct forebears but also by observing mores of modern hunting groups extant in today's tropical forests. Besides the minimally essential foodstuffs available to sustain the australopithecines, small amounts of additional energy were likely used both to gather food and prepare it for consumption. For example, early domestication and subsequent use of dogs would have aided the hunt for food, but of course the dogs also need nourishment. Fire useful in the hunt as well as in the preparation of some foods would have also utilized more energy; possibly as long ago as 165 ky, not only for cooking but also for heat-treating stones to make better tools [196], the exploitation of energy would have roughly doubled $\Phi_m$ to ~40,000 erg/s/g for slightly heavier, archaic *H. sapiens*. Ample evidence exists that even earlier hominids, notably *H. erectus*, used pits for roasting animals and perhaps even for drying food prior to its preservation and storage to guard against lean periods. Fire also allowed the preparation of certain vegetables known to have been then widely consumed, such as yams that require washing, slicing, and leaching with hot water to remove alkaloid poisons. However, claims that hunter-gatherers used more energy than modern humans are dubious, caused by overestimates of the former and underestimates of the latter [197]; deliberate burning of land unhelpfully dissipates energy as waste heat and polluting smoke directly into the air thereby performing no real work for aborigines, although such widescale destruction does usefully clear land for more efficient crop plantings that later bolster consumption. (Reference [22] further discusses waste heat that, even today, provides no beneficial energy to humankind while degrading surrounding environments.) Recent anthropological field studies of cultural evolutionary strategies based on energy use are consistent with hominid $\Phi_m$ values used in this paper [189]. To what extent hunter-gatherers merely used fire when and where available, in contrast to actually possessing it or controlling it, is unknown—but fire does grant, at least in some small way, an energy supplement to the basic metabolic budget of early humans.

Agriculturists ~10 kya not only used fire but also clearly controlled it, constructed irrigation ditches and terraced fields, probably deployed rudimentary windmills and watermills, and engaged draft animals to plow fields more deeply and extensively (such animals typically delivering ~600 W of power, compared to human exertion averaging 75 W)—all with the intent of increasing crop productivity. Anthropologists have documented such advances for more recent, if still prehistoric, times, especially where remains of fully domesticated varieties of plants and animals are present in archaeological contexts. Many locales independently pioneered agriculture including, for example, southwest Asia (~9 kya, or ~7000 y BCE), the Middle East and Mediterranean (~8 kya), and Meso-America (~7 kya), although it may well have begun in western Asia where collections of wild grains are found ~11 kya among nomadic tribes who were still at the time hunter-gatherers. Later domestication allowed human societies to actively alter the genetic composition of organisms by breeding (*i.e.*, replacing traditional biological selection with human-directed cultural selection, mostly by trial and error in the absence of any knowledge of genes), thereby cultivating plants such as maize (now 7 times the size of its original, undomesticated cobs) and sugarcane (now much more efficient than its natural strain). The poverty of energy apparently limited cultural development, yet with the onset of agriculture and the use of trained animals ~10 kya, the equivalent energy available to individual *H. sapiens* (assumed here to be a 50-kg body) increased $\Phi_m$ to ~12,000 kcal/day, or ~$10^5$ erg/s/g; in turn, these would have easily doubled with the invention of advanced farming techniques and the invention of metal and pottery manufacturing a few millennia ago. (Today, the most intensive agricultural methods yield as much as 40,000 kcal/day/person.) During this energy-enhanced Neolithic Revolution, ecosystems shifted from food collection gathered in the wild to food production by deliberately managed means, and the results a few thousand years later included the advent of local cities, professional warriors, regional alliances, and ultimately nation-states. Agriculture's greatest achievement was to feed the growing human population, which rose from ~170 million people ~2 kya (1 CE) to ~450 million some 500 ya and to ~900 million about 200 ya [198]. Underlying all this cultural advancement was greater energy usage per unit mass at each and every step of the way.

Industrialists of a couple centuries ago learned to use energy to power machines in their homes and shops, thereby causing huge demand for fossil fuels and hydropower, which in turn transformed the production of goods, agriculture, transportation, and communications. The burning of coal at the start of the energy-driven Industrial Revolution afforded each member of a young, mechanistic society (especially in Britain, Germany, and the United States) a great deal more energy for use in daily, societal activities. As human population rose greatly by ~5 billion people since 1800 CE, reaching ~6 billion by the year 2000, per capita energy usage also increased—in fact, well exceeded the energy contained in the food that people physically consumed or even produced. Total energy utilized during this period climbed dramatically and globally, much more so than when our earlier ancestors mastered pragmatic fire or invented solar-based agriculture. Comparing the prior agricultural age with the current fossil-fuel-driven industrial age, per capita energy usage likely tripled; this agrees with the ~1.6 kW per capita (~$3\times10^5$ erg/s/g) usage reported today for most industrial nations, including food, fuel, electricity *etc*. [199, 200]. Across the world currently, each citizen averages $5\times10^5$ erg/s/g, which is roughly an order of magnitude more than our hunter-gatherer forebears. Again, as with estimates of $\Phi_m$ for galaxies, stars, plants, and animals discussed above, this is an average value within a range of variations, since residents of advanced, OECD (Organization for Economic Cooperation and Development) countries, such as those in Europe and N. America, use several times more, whereas developing (non-OECD) countries, such as China, India and all of Africa, use several times less. For example, per capita expenditure of energy now averages 2.6 kW globally, yet varies regionally from ~0.5 kW for Africa to ~4.5 kW for Europe and to ~12 kW for N. America [201]. The result, ecologically, is that the stored energy of fossil hydrocarbons has been added to the daily energy arriving from the Sun (and more recently that of terrestrial nuclear energy as well), all of which are employed by human societies in various ways to access more resources and yield yet more productivity as well as to change the very fabric of our earthly environment. Such unprecedented application of energy to produce goods, services, and knowledge (which, in turn, furthers the acquisition of still more energy) has also taken a toll on that environment. Regardless of all else, the 2$^{nd}$ law of thermodynamics demands that as any system complexifies—even a human social system—its surrounding environment necessarily degrades.

Technologists, also known as consumer-traders in today's world, represent the most highly developed and energy-intensive, yet wasteful, part of contemporary society, displaying during the past half-century large electricity and transportation allocations throughout their energy budgets. Perhaps we are creating a Digital Revolution, but its root cause is still energy based. Distinguished from industrialists, technologists employ an energy rate density (>$10^6$ erg/s/g) that is several times greater than that of traditional commercial society (perhaps epitomized by astronaut-elites who individually enjoy energy shares of ~$10^7$ erg/s/g while orbiting aboard the International Space Station, or an equivalent per capita energy use of more than 1 million kcal/day, which is fully ~500 times more than each of us actually consumes as food daily). Symbolized by the most heavily energy-using countries such as the United States, Canada, Bahrain, and Qatar, technological societies have distinctly higher $\Phi_m$ values than the average global citizen on Earth today or even than those living in the developed countries of Europe. A single example of such energetic excess will suffice: With coordinated power generation and widespread distribution systems boosting the effective daily energy used, the per-citizen expenditure in *all* countries averaged 55,000 kcal by 1970, or ~$5 \times 10^5$ erg/s/g; now, early in the 21$^{st}$ century, with ~25% of the world's total power exploited by only 5% of the world's population mostly living in the U.S., this one country averages $2 \times 10^6$ erg/s/g (which amounts to ~12.5 kW for each U.S. citizen, compared to ~2.6 kW per person globally). Thus, modern high-tech conveniences, from automobiles, airplanes, and centralized heating/cooling devices to a wide variety of energy aids enhancing our digital society (including wired homes, networked businesses, and consumer electronics of all sorts), empower today's individuals well beyond their daily food intake [202, 203]. All these energy budgets are still rising—in both absolute terms as well as per capita accounts.

Figure 8 plots the increase of $\Phi_m$ as culture advanced and humanity complexified in relatively recent times (see also Table 1 of [22]). Note how industrialists of hundreds of years ago had higher energy rate densities than agriculturists or hunter-gatherers of thousands of years ago, and, in turn, energy affluent western society still higher values today. Here, social progress, expressed in terms of per capita energy usage, is graphically traced for a variety of human-related strides among our recent hominid ancestors. Note how the rise has been truly dramatic in very recent times as our civilization became, as it remains now, so heavily wedded to energy for its health, wealth, and security.

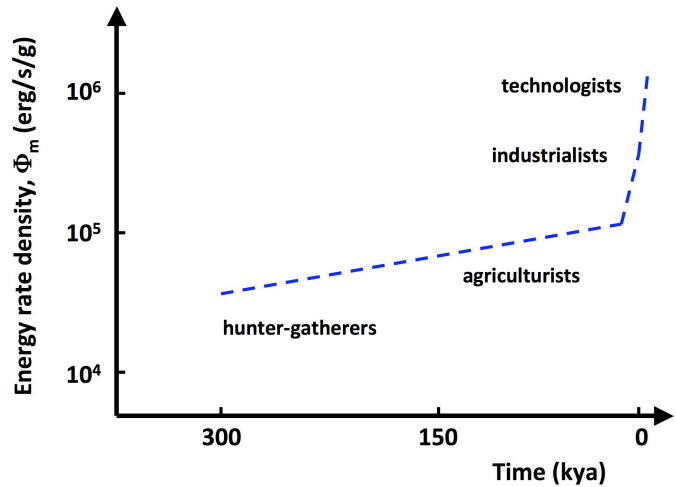

FIGURE 8: The temporal dependence of energy rate density for human society as plotted here pertains only to the topmost part of the master graph in Figure 2, and thus concerns less than the past million years or <0.01% of all of cosmic history. This graph serves to illustrate the advance of (per capita) energy usage by some of our hominid ancestral groups during the cultural-evolutionary phase of cosmic evolution. On a linear temporal scale as plotted here, that rise is approximately exponential in very recent times as civilization has become heavily dependent upon energy for its continued well-being.

*4.7.3 Technology Evolving.* Foremost among the advances that helped make us cultured, technological beings were the invention and utilization of tools, which require energy to build and operate, once again decreasing entropy within those social systems using them while increasing it in their wider environments beyond. Thermodynamic terminology may be unfamiliar to cultural anthropologists or big historians, but the primary energy-based processes governing the cultural evolution of technological society are much the same, albeit measurably more complex, as for the evolution of stars, galaxies, and life itself.

Anthropocentrism need not enter the cosmic-evolutionary narrative here. Just because powered devices are structured by, or perform functions for, humanity does not mean that their cultural complexity need be analyzed any differently from other forms of complexity (*cf.*, Sec. 5.9). Rather, it seems reasonable and consistent, backed by the quantitative research of this paper and especially [22], that humans and their machines, among all other complex systems and across the history of time to date, are merely members of a continuum of rising complexity, from the origin of the Universe to the present. History repeatedly shows that we like to regard ourselves and our accomplishments as special, yet, even in the unlikely event that we are alone in the Universe, it is still probable that evolved humans and our built machines are no different, at any kind of basic level, than any other complex system in our amply endowed Universe.

*4.7.4. Energy Rate Density for Machines.* One of the most prominent cultural icons in today's world is the automobile, and not just for developed countries where citizens can afford this kind of machine transport. Motor vehicles are now ubiquitous across planet Earth, for better or worse archetypical symbols of technological innovation in our modern society. Evaluating machines in the same energy-based way consistently applied throughout this research program, we can compute a value of $\Phi_m$ for an average-sized automobile, whose typical properties are ~1.6 tons of mass and ~$10^6$ kcal of gasoline consumption per day; the answer, $\Phi_m \approx 10^6$ erg/s/g (assuming a few hours of operation daily), is likely to range higher or lower by several factors owing to variations among vehicle types, fuel grades, and driving times; this average value approximates that expected for a cultural invention of considerable magnitude—indeed, for what some still claim is the epitome of American industry. Put another way to illustrate not only high degree of complexity but also evolutionary trends, and using numbers provided for the past quarter-century by the U.S. Highway Traffic Safety Administration [204], the horsepower-to-weight ratio (in English units of hp/100 lb) of American passenger cars has increased steadily from 3.7 in 1978 to 4.1 in 1988 to 5.1 in 1998 to 5.5 when last compiled in 2004; converted to the units of $\Phi_m$ used consistently throughout this paper, these values equal 6.1, 6.7, 8.4, and 9.1, all times $10^5$ erg/s/g respectively. (By comparison, a literal draft horse's power density equals ~745

W/800 kg, or ~$10^4$ erg/s/g, a value appropriately within the midst of the mammalian range, as noted in Sec. 4.5 on animals above). Not only in and of themselves but also when compared to less powerful and often heavier autos of >50 ya (whose $\Phi_m$ values averaged less than half those above), the span of these numbers confirms once again the general correlation of $\Phi_m$ with complexity. No one can deny that modern automobiles, with their electronic fuel injectors, computer-controlled turbochargers, and a multitude of dashboard gadgets are more complicated than Ford's "Model-T" of nearly a century ago—and that more energy is expended per unit mass to drive them.

The evolution-$\Phi_m$-complexity correlation hypothesized here can be more closely probed by tracing the changes in internal combustion engines that power automobiles among many other machines such as gas turbines that propel aircraft [58, 205]—all notable examples of technological innovation during the power-greedy 20$^{th}$ century. To be sure, the brief history of machines can be cast in evolutionary terms, replete with branching, phylogeny, and extinctions that are strikingly similar to billions of years of biological evolution—though here, cultural change is less Darwinian than Lamarckian, hence quicker too. Energy remains a key facilitator of these cultural evolutionary trends, reordering social systems much like physical and biological systems from the simple to the complex, as engineering improvement and customer selection over generations of products made machines more elaborate and efficient. For example, the pioneering 4-stroke, coal-fired Otto engine of 1878 had a $\Phi_m$ value (~$4 \times 10^4$ erg/s/g) that surpassed earlier steam engines, but it too was quickly bettered by the single-cylinder, gasoline-fired Daimler engine of 1899 (~$2.2 \times 10^5$ erg/s/g), more than a billion of which have been installed to date in cars, trucks, planes, boats, lawnmowers, *etc.*, thereby acting as a signature force in the world's economy for more than a century. Today's mass-produced automobiles, as noted in the previous paragraph, average several times the $\Phi_m$ value of the early Daimler engine, and some racing cars (akin to temporarily active galaxies or metabolically enriched race horses and Olympic sprinters) can reach an order of magnitude higher. Among aircraft, the Wright brothers' 1903 homemade piston engine (~$10^6$ erg/s/g) was superseded by the Liberty engines of World War I (~$7.5 \times 10^6$ erg/s/g) and then by the Whittle-von Ohain gas turbines of World War II (~$10^7$ erg/s/g). Boeing's 707 airliner inaugurated intercontinental jet travel in 1959 when $\Phi_m$ reached ~$2.3 \times 10^7$ erg/s/g, and civilian aviation evolved into perhaps the premier means of global mass transport with today's 747-400 jumbo-jet whose engines generate up to 110 MW to power this 180-ton craft to just below supersonic velocity (Mach 0.9) with $\Phi_m \approx 2.7 \times 10^7$ erg/s/g.

The rise in cultural $\Phi_m$ values can be traced particularly well over several generations of jet-powered fighter aircraft of the U.S. Air Force, further testifying to the ever-increasing complexity of these sophisticated, supersonic machines. (Note that engine thrust must be converted to power, and for unarmed military jets operating nominally without afterburners 1 N ≈ 500 W, for which $\Phi_m$ values then relate to thrust-to-weight ratios). First-generation subsonic aircraft of the late 1940s, such as the F-86 Sabre, gave way to 2nd-generation jets including the F-105 Thunderchief and then to the 3rd-generation F-4 Phantom of the 1960s and 70s, reaching the current state-of-the-art supersonic F-15 Eagle now widely deployed by many western nations; 5th-generation F-35 Lightning aircraft will soon become operational. (Fighter F-number designations do not rank sequentially since many aircraft that are designed never get built and many of those built get heavily redesigned.) These aircraft not only have higher values of $\Phi_m$ than earlier-era machines, but those energy rate densities also progressively rose for each of the 5 generations of aircraft R&D during the past half century—2.6, 4.7, 5.7, 6.1, and 8.2, all times $10^7$ erg/s/g respectively, and all approximations for their static engine ratings [206].

This discussion of the rise of machines is extended in [22] to include the origin and evolution of computers, which also effectively exhibit increases of $\Phi_m$ with the advancing evolution of computer complexity during the past few human generations. In all, the quantitative assessments of machines provide a remarkably good reality check of this admittedly unorthodox, thermodynamic interpretation of cultural evolution.

Figure 9 depicts several of the above-derived values of $\Phi_m$ for culturally devised machines. Engines are only one of a multitude of technical devices invented, improved, and now deployed by humankind on Earth; many other cultural advances could be similarly appraised and most would display comparably high values of $\Phi_m$. This graph illustrates for today's technologically sophisticated society, much as for so many other complex systems examined throughout the cosmic-evolutionary scenario, how energy rate density parallels the rise of complexity in time. As with previous surveys of many complex systems, built machines are merely another example (albeit among the latest on Earth) of rising complexity with the advance of evolution writ large.

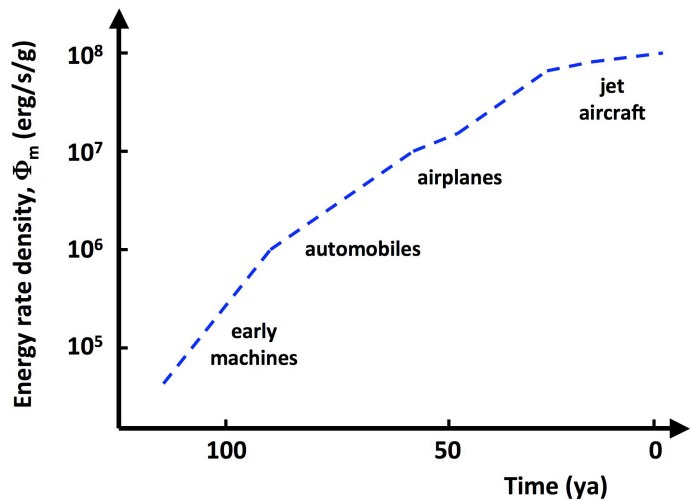

FIGURE 9: The complexity of machines, expressed in terms of $\Phi_m$, rises to illustrate increased utilization of power density by human-built devices during the cultural-evolutionary phase of cosmic evolution. That rise has been dramatic (shown here over 3 orders of magnitude) within only the past few human generations as technological civilization has become increasingly dependent upon energy. Note that the timescale for this graph is much, much briefer than any of the previous figures—here, roughly the past century of natural history—so it represents only a miniscule part atop the master curve graphed in Figure 2.

*4.7.5. Summary for Society and Its Machines.* Human society and its invented machines are among the most energy-rich systems with $\Phi_m > 10^5$ erg/s/g, hence plausibly the most complex systems known in the Universe. All of the culturally increasing $\Phi_m$ values computed here—whether slow and ancestral such as for controlled fire and tilled land, or fast and modern as for powered engines and programmed computers in today's global economy—relate to evolutionary events in which energy flow and cultural selection played significant roles:

- from agriculturists ~10 kya ($\Phi_m \approx 10^5$ erg/s/g)
- to industrialists some two centuries ago (~$3\times10^5$)
- to automobiles and computers decades ago (~$10^6$)
- to technologists in highly developed countries today (~$2\times10^6$)
- to computer-controlled jet aircraft of today (~$10^7$).

## 5. Clarification of Key Concepts

Confusion and misinterpretation often arise when carefully constructed journal articles go unread amid today's harried world of hasty e-mails, biased internet blogs, and unrefereed papers in some open-access outlets. Needless anxieties also result when scientists write for non-science audiences (and likely conversely)—and my experiences with big historians are no different. Natural scientists often cringe at many of the soft pronouncements of humanistic and social scholars, while big historians find challenging the quantitative propensity of hard science.

This section attempts to clarify some subtle concepts and quantitative accounts pertinent to the uncommonly wide array of sciences undergirding big history—in that way, at least making clear my stance on several issues shaping the scientific basis of this newly profound academic interdiscipline. I shall also note some criticisms, and also self-criticisms, of my work, identifying several areas where more research is needed to further build a professional foundation for big history using the best available empirical evidence and scholarly methods.

*5.1. Self-organization.* Self-assembly, self-organization, and self-ordering do not exist in Nature. Dynamical processes in which "interacting bodies are autonomously driven into ordered structures" always involve energy [207, 208]. Energy is inevitably engaged in any transaction that forms structural and functional patterns; the origin, maintenance, evolution, and fate of all systems are infused with energy. Popular terms that imply self-alteration, or even self-sustenance, of complex systems are inaccurate descriptions of real, material phenomena; they often mislead non-experts who regard complex systems as anomalous and confuse non-scientists who think that systems emerge spontaneously or change magically all by themselves. Influential scientific organizations (*e.g.*, NASA) even define life as a "self-sustaining chemical system capable of Darwinian evolution" [209]; such definitions convey mainly that life displays metabolic properties, yet all metabolisms run on external energy, hence often create misconceptions in science education. Renowned colleagues regularly yet vaguely assert self-organization as the basis for life's structure and function, often bolstered with elegant mathematics yet devoid of empirical data justifying the transcendent leap to *self*-organization from their otherwise reasonable stance that physical laws govern chemistry, biology, and the process of evolution itself [27, 40, 210, 211]. No unambiguous evidence exists for any event in Nature occurring spontaneously, alone, or without energy exchange; energy of some type, at some level, and for some time seems always involved in any material change.

*5.2. Non-equilibrium.* Equilibrium represents a minimized energy state for any system of any geometrical configuration; if complex systems experience no appreciable agents acting on them, they will naturally relax to a state of minimum energy. However, robust energy is integrally involved in the origin and evolution of real physical, biological, and cultural systems—regular inflows of energy, which literally drives (*i.e.*, forces) them away from the disordered equilibrium of isolated systems that have no external input. In particular, energy flows provide a physical basis for biological life, allowing life to sustain excursions far from equilibrium—and for cultural activities as well, maintaining cities, societies, and civilization itself in dynamical steady states of order and organization while temporarily imbalanced. Humans, too, are well removed from equilibrium, and provided we maintain an energy intake within an optimal range our bodily structures and functions remain viable; once we stop eating, we do eventually achieve a balanced equilibrium at death. The impermanent cells of our human bodies renew completely every ~7 years, those of our skin in a few weeks and some in our gut in a few days; likewise, stars constantly make heavier nuclei, galaxies regularly form new stars, and society frequently innovates, reinventing ways to organize humanity and build gadgets. If Nature were actually equilibrated (and thus its entropy maximized), all life-forms would be dead, and stars and galaxies would not exist. Many leading scholars distort the language of these scientific concepts: traditional clinicians practicing Freudian psychology wrongly equate good health with equilibrium, even while regarding mental urges and instincts as gusts of energy swirling through the brain [212]; orthodox economists model goods exchange in the marketplace as closed systems that are supply-demand equilibrated, even while realizing that cities can survive only as open systems with food and fuel flowing in while products and wastes flow out [213]; even pioneering big historians who helped establish their new subject as a legitimate field of scholarship, indeed gave it its name, declared that stars, Earth, and humanity itself are all safely ensconced within equilibrium regimes, which is fundamentally at odds with the findings of natural science [214, 215]. In equilibrium, time is irrelevant and thermostatics pertain yet explain little; by contrast, real systems obey *non*-equilibrium thermo*dynamics* where time is of the essence, change is ubiquitous, and energy is centrally engaged.

*5.3. Optimization.* Life seems to function optimally within certain boundary conditions and not surprisingly also has an optimal range of normalized energy flow; so do all other complex systems. The vast majority of $\Phi_m$ values for both plants and animals fit neatly (with some variation and overlap) between inanimate physical systems having lower $\Phi_m$ and more advanced cultural systems having higher $\Phi_m$. That the trend of increasing $\Phi_m$ values with the evolution of living systems is imperfect should not deter us, for the great diversity of animals often display wide physiological adaptations to extreme environments, and in any case no useful investigation can proceed if it must justify every rarity or outlier. In fact, it might be those same variations in $\Phi_m$ that grant life-forms opportunities to advance and further complexify; without variation, life would likely stagnate, as would all complex systems. Much is also the case for stars that need certain threshold energies to ignite fusion (in protostars) yet not so much energy as to explode violently (in supernovae). Optimality is likely favored in any system's use of energy—not too little as to starve it, yet not too much as to destroy it. Societies, machines, cities, among other cultural systems also display energy flows within certain optimal ranges—different ranges for different systems of different masses—and if those systems acquire too little or too much energy they abort, reverting to their simpler selves. Thus, my hypothesis addresses both growth of complexity and return to simplicity, as well as stipulating those conditions when either outcome is favored. Some complex systems do indeed run afoul of energy flows outside their optimality ranges, thereby devolving to simpler status of which there are many examples in Nature, and not just in biology—*e.g.*, white dwarf stars become homogenized near termination, cavefish lose their eyesight while retreating to darker niches, and cities go bankrupt while unable to manage sufficient energy flows for their residents. None of these failures are exceptions to cosmic evolution writ large; rather, such troubled systems often persist for some time in reduced complex states before collapsing outright and eventually becoming extinct [216]. By contrast, successful complex systems seem neither fine-tuned nor perfectly built, nor do they exhibit maximum energy flows or minimum entropy states. Rather, optimization is a constraining feature for a bracketed range of maximum and minimum values of $\Phi_m$, above and below which, respectively, a system cannot function—an empirical finding that I have stressed for many years in many peer-reviewed publications (*e.g.*, [8, 9, 15, 217, 218]). More recently, big historians have reappropriated the key idea of optimization under the guise of "Goldilocks conditions" or "Goldilocks circumstances" [3, 5, 219], but there is no need to relabel the scientifically based concept of energy-optimization by appealing to humanistically inspired fairytales. Boundary conditions that are not too hot and not too cold, or physical dynamics that are neither too fast nor too slow, *etc.*, rather "just right" to create and sustain complex systems, are synonymous with optimal energy ranges (also just right) that have long been employed by natural scientists; some astronomers, for example, cast Earth's habitability in Goldilocks-laden descriptions—if Earth were nearer to or farther from the Sun, or if our atmosphere were thicker or thinner, or if it were abundant in this or that composition, then Earth might be unsuitable for life—yet these are hardly more than flamboyant restatements that only certain amounts of energy are available at Earth's surface, and that if conditions were different we might not be here. Environmental conditions per se are not an underlying reason for complexification; energy flows through systems likely are; energy is the cause, complexity the effect. There is no need to reinvent soft terms that invoke myth or fantasy, yet which cheapen the hard science describing such complex systems; there is nothing intractable here—although big historians may think so when relating the history of humans and their cultural inventions, which some of them apparently regard as special or separate from other systems in the Universe (*cf.*, Sec. 5.9). Big history is not a recounting of imagined fables and magical powers; rather, it is a wonderful new way to scientifically chronicle all of history, from big bang to humankind, without assuaging this grand narrative with equivocal terms and fictitious notions that sow doubt and misconception, yet skirt serious understanding of how material systems might emerge, mature, and terminate. If big historians want their magnificent story to be empirically based, they ought to accept some objective, quantitative terms; linguistic attempts to soften hard science will likely lead to subjective, qualitative confusion and needless controversy.

*5.4. Predictions.* Evolution is not a predictive science. No one knows specifically where the master curve of rising complexity (Figure 2) is headed, other than presumably toward greater complexity in an expanding Universe. Given that random chance is an intrinsic part of evolution on any scale, at any time, and for any system, there will always be a strand of uncertainty in the outcome of any change. However, as Nature selects for or against system viability, determinism is also a vital part of the action. The two—chance and necessity—work in tandem, comprising the process of differential *natural* selection (and not only for biological systems, rather likely for every complex system—*cf.*, Sec. 5.6), which acts as a ruthless editor or pruning device to delete those systems unable to command energy in optimal ways. That is why I have always preferred "non-random elimination" as a more appropriate description for natural selection broadly applied to all complex systems [220]. No one knows, or probably ever will, the proportions of each in any given transaction; some [221-223] favor chance, others [32, 224] necessity. That same inexact mixture of randomness and determinism is also why realistic outcomes of most changes will never be precisely predictable, but will remain process-dependent and undetailed; all systems that obey non-linear dynamics preclude predictions far into the future [225]. Even so, attempts to describe the grand scope of evolution quantitatively do display some general trends among a rich compendium of available data. As best can be determined presently, a perpetual advance toward ever-richness, diversity, and complexity, specific outcomes of which cannot be foreseen, may be the ultimate fate of the Universe.

*5.5. Structure and Function.* Complex systems exhibit both form and function; the former is a system's structure, the latter is what it does (without any metaphysical "plan" or "purpose" implied). Structure seems prerequisite for any system's properties, and an essential feature for any system to function; the structure of a protein, for example, is crucial to its function, and if it has the wrong structure it will malfunction. Some simple structures have little or no function and are therefore not sustained systems; structures acquire and store energy, but only function can express energy. A rock in the backyard exemplifies a system with limited structure yet no real function, which is why most rocks are not very complex entities and have low values of $\Phi_m$. Typical rocks, now part of a cooling Earth (*cf.*, Sec. 4.3), comprise modest inhomogeneous material composition—the result of physical evolution during more formative periods in our planet's history when energy flows through a differentiating mantle were greater than today; only minute amounts of radioactive decay now keeps such rocks, which alone show no function, from having 0 erg/s/g. A raw egg is another familiar example of a relatively simple system, even when rich in organic matter, displaying little structure and no actual function. If an egg is smashed on the pavement, any organized structure it had within its shell is irreversibly destroyed—a clear case of a system exceeding its optimal range in energy rate density. But if an egg is more moderately (and optimally) energized, such as during sustained boiling, it effectively utilizes the acquired heat to become somewhat more structurally complex as proteins within unfold and aggregate, which is evident when its shell is pealed away to reveal its yoke-white inner order. Yet, even when hard-boiled and mildly structured, an egg has no evident function and its value of $\Phi_m$ is small. More organized systems, especially those experiencing biological and cultural evolution that typically have higher $\Phi_m$, also have greater amounts of structural intricacy in addition to enhanced functionality. Animals (*cf.*, Sec. 4.5), which are both considerably structured and actively functioning (even while resting but not when dead), exhibit values of $\Phi_m$ that are orders of magnitude larger than any inanimate system often because both structure and function contribute; birds have amplified values of $\Phi_m$ as computed above largely because of their especially impressive function of flying three-dimensionally. Machines (*cf.*, Sec. 4.7) also display both types of complexity though not always sustained; computers have much structural complexity (energy stored) and a high value of $\Phi_m$ when functioning (energy expressed), yet when turned off have no function and hence no energy rate density; mousetraps likewise have some structure but no frequent function—until such time when the lever is tripped, the stored energy released, and the mouse terminated. And as for brains (*cf.*, Sec. 4.6), if their neuronal meat is the structure, then their conscious mind is the function, and probably nothing much more or mystical than that. Again, structure seems precedent, fundamental, perhaps even a precondition for viable function. Structure can exist without function, yet not conversely, thus the aesthetic cliché "form follows function" is probably reversed for most complex systems in Nature, much in accord with Darwinism generally and with apologies to architects everywhere. Part of any system's $\Phi_m$ value derives from structure and part from function, the two likely being multiplicative more than additive; apportioning relative contributions to total system complexity is non-trivial, indeed likely impossible currently. Unraveling the proportions of complexity attributable to each form and function will someday help reveal the devilish details needed to quantitatively explain the full nature of systems complexity.

*5.6. Natural Selection and Adaptation.* The word "evolution" should not be restricted to biology alone; a broad interpretation of this term generally applies to all complex systems, living or not, thus the subject of cosmic evolution includes physical, biological, and cultural evolution. Likewise, the process of "selection" can be considered generally, as it *naturally* affects complex systems throughout Nature; hence, natural selection applies not merely to living systems but to all systems that naturally experience physical, biological and cultural selection. Thus, to be clear as I see it, biological evolution occurs by means of biological selection (*i.e.*, neo-Darwinism)—yet all systems, including those that are inanimate and cultured, evolve by means of comparable selection. This is not to claim that either physical or social systems change in the same specific ways as do biological systems. System functionality and genetic inheritance—two factors above and beyond system structure—enhance complexity among biological systems that are clearly living compared to physical systems that are clearly not. For animate systems, energy is fuel for change, helping at least in part to select systems able to utilize increased power densities, while driving others to destruction and extinction—all in accord with neo-Darwinism's widely accepted modern synthesis. Nothing in this paper disputes neo-Darwinism; the facts of biological evolution are unassailable even if the mechanism by which it works is still unresolved. As proposed here, energy flow, provided it is optimally favored by a mutated living system's altered genome, is envisioned to aid biological selection; energy itself conceivably acts as a central means by which biology's evolutionary mechanism works. Energy flow and biological selection likely operate together as life-forms mutate—the former utilized by those systems advantageously suited to their randomly changing environments, and the latter non-randomly eliminating those unable to do so (*cf.*, Sec. 5.3). Biological selection thereby shapes phenotypes near an adaptive optimum, yet in reality traits often vacillate around that optimum as viability and variability go hand in hand (*cf.*, Sec. 5.8). Likewise, for cultural systems, much of social advancement is aided and abetted by acquired knowledge accumulated from one generation to the next, including client selection, rejection, and adaptation—a Lamarckian more than Darwinian process. Cultural inventiveness enabled our immediate ancestors to evade some environmental limitations, such as hunting and cooking that allowed them to adopt a diet different from that of the australopithecines, while clothing and housing permitted them to colonize both drier and colder regions of planet Earth. That's not biological (Darwinian) selection, but it is cultural selection; any complex system is *naturally* selected or rejected by means of interactions with its environment—and that includes, and might be dominated by, energy flows in the area. For all biological and cultural systems, if the energy acquired, stored, and expressed is optimum, then those systems survive, prosper, and evolve; if it's not, they are deterministically selected out of existence. Physical systems, too, are not much different, even if physical selection (which is also part of natural selection) operates less robustly and adaptation reduces to simple adjustments [16 notably Sec. 5; 226]). To be sure, selection and adaptation are not exclusively within the purview of biology; inanimate systems also experience these twin agents of change, albeit in rudimentary ways that preserve the fittest variants between any complex system and its surrounding environment. Examples abound: Pre-biological molecules bathed in energy were selected in soupy seas to become the building blocks of life; certain kinds of amino-acid bonding were favored while others were excluded, implying that the evolutionary steps toward life yielded new states more thermodynamically stable than their precursor molecules. Crystal growth among many other non-living systems (such as clays) also displays simplified selection; ice crystals grow and slightly complexify when water molecules collide and stick (much as do snowflakes, once thought to be perfectly symmetrical 6-sided beauties, yet which are mostly irregularly ordered conglomerates displaying great morphological variation), and although the initial molecular encounters are entirely random the resulting electromagnetic forces that guide them into favorable surface positions are not. Even stars exhibit crude adaptation and selection: Our Sun adjusts to changing conditions by naturally increasing its internal chemical and thermal gradients during fusion, yet it will not be selected by Nature to endure beyond a carbon-oxygen mix largely because its energy flow will fail to reach the critical threshold needed for the natural emergence of greater complexity. Not just developmentally in a single stellar generation while passing from "birth" to "death" but also over multiple generations, stars are widely acknowledged to physically evolve; much akin to changes within populations of plants and animals over many generations of life-forms, the most massive stars selected to endure the increased fires needed to make heavier nuclei are in fact the very same stars that often create new populations of stars, which in turn do display increased $\Phi_m$ values as $2^{nd}$, $3^{rd}$, and $N^{th}$-generation stars emerge from interstellar debris—none of which means that stars are alive or evolve biologically, a frequent though invalid criticism. All things considered, natural selection is a universal phenomenon dictated by not mere chance, nor even by only chance or necessity; rather natural selection within and among all complex systems engages both chance *and* necessity. Nothing in Nature seems black or white, rather more like messy shades of gray throughout.

*5.7. Emergence.* Tenably, energy drives systems beyond equilibrium while selection aids the emergence of greater complexity for those systems able to manage the increased energy flow per unit mass. In other words, normalized energy flow might itself be the trait most often selected by successful systems of the same kind; if so, emergence becomes technically synonymous with creativity. (See [227] for a brief review of the slippery concept of emergence.) Perhaps not as mysterious or magical as some complexity scientists imply, emergence might be hardly more than the straightforward outcome of ways that energy naturally and hierarchically enriches system structure *and functionality.* Clocks and phones tell time, birds and aircraft fly high, ants and cities network; all these systems and so many others admittedly demonstrate properties not seen among pre-existing, lower-level components of less complex systems. However, emergence need not be anything more than novel properties gained by virtue of systems' increased degree of complexity; new system properties likely emerge when favored systems naturally evolve across critical thresholds marking higher degrees of complexity [64, 208]. Complex systems obey non-linear dynamics that commonly exhibit phase-transition bifurcations—sudden changes in behavior as some parameter of the system alters, such as the famous case of rapid onset of convection rolls in a fluid heated from below once a temperature gradient exceeds some threshold. An ecology is another good example: Small fluctuations in diverse ecological systems are not often canceled by some other change, thus destroying any "balance of Nature," which actually exists nowhere within or among realistic, non-equilibrated systems. Rather, energy acting on such fluctuations can sometimes cause them to grow dramatically via positive feedback into something yet more complex—again, both structurally *and* functionally; energy flows exceeding a critical threshold can drive a system far beyond equilibrium, where selection can, if energy is optimized for that system, aid the emergence of demonstrably new properties—an underlying physical process that probably governs what other researchers call self-assembly (*cf.,* Sec. 5.1). This is evidently why most complex systems fail: they are often more challenged to optimize, hence more fragile to sustain, than are simpler systems and their risks tend to accumulate rather than cancel. It is the simpler systems that usually survive best; there are many more relatively simple dwarf stars in the sky or microbes in our bodies than advanced civilizations on habitable planets. None of these predilections invoke reductionism as much as holism, the latter a non-metaphysical expression of bottom-up systems analysis; reduction and holism, like chance and necessity, also have their shades of gray. Yet "more" complexity need not mean "different" [228], rather authentically more, as lower-level symmetries break, causing not only existing systems to complexify but also new systems to form capable of utilizing increased energy rate density. Sometimes loosely termed "novelty by combination," emergence is a genuinely holistic phenomenon at work everywhere in Nature, much as Aristotle long ago posited "the whole to be something over and above its parts," from atomic physics to organic biology to human culture. Among prominent examples, liquid water's covalent-bonded properties are not deducible from the elemental properties of its components, O and H, both of which are colorless gases, one of them explosive; nor is NaCl, which we enjoy as ionic-bonded table salt despite one of its atoms being a toxic poison. Likewise, living organisms are more than magnified manifestations of their constituent molecules; evolutionary biologists usually study entire cells and complete organs, not merely individual molecules and genes, yet all are required for full understanding. Society is also more than a mere assembly of its member citizens, as opined earlier (*cf.*, Sec. 4.7) when $\Phi_m$ for civilization was numerically shown to exceed by roughly a factor ten that for the individual humans comprising it. To my mind, emergence might simply be a natural way that favored systems complexify, maturing additional, yet not necessarily different, "intricacy, complication, variety or involvement" (my definition, Sec. 2) with the march of time; incremental quantitative changes caused fundamentally yet partially by energy flows conceivably and unpredictably lead to qualitative novelty "over and above" a system's many varied, interacting, reducible parts. If correct, life itself and its consequent behavior are hardly more than an energetic driving of organic molecules out of equilibrium (*cf.*, Sec. 5.2) sufficient to create emergent structures with functions as complex as those of living systems; life needs no mystical properties any more than it needs *élan vital*, which faded away with improved insight after decades of struggles to decipher it. Perhaps it's too much for one paper to challenge *both* of the cherished concepts of self-organization and enigmatic emergence so central to orthodox complexity science. Critics will likely judge my attitude as an abandonment of holism and a retreat to reductionism, which it is not; complex systems can indeed manifest more than their whole yet less complex parts. Rather, I regard my considered temperament as a promising way to evade opaque mysticism while promoting quantitative synthesis throughout natural history.

*5.8. Outliers.* Complex systems that are struggling, collapsing, or otherwise have abnormally high or low values of $\Phi_m$—whether aged stars, endangered species or troubled societies—are often considered exceptions of cosmic-evolutionary cosmology. However they are not; they are among many systems that naturally display abnormality while evolution non-randomly selects winners and eliminates losers. Nature is rich in outliers, indeed they are sometimes beneficial for diverse, changing, complex systems; without variation evolution would not produce novelty and creativity seen throughout the Universe. Our own species has plenty of variation, for instance, with its obese and malnourished members or tall footballers and short jockeys; dwarf and giant stars, dark and active galaxies, prosperous and broken cities—variations are everywhere. Outliers' anomalous values of $\Phi_m$ cannot be dismissed or explained away, as they are often genuine variations within a normal (Gaussian) distribution around some mean value. Critics complain that their favorite bird, jellyfish, or gadget do not lie exactly on the appropriate curve in Figures 3-9, which is often true; yet, given the inherent diversity of systems, we should realistically expect to find only small minorities of complex systems (if any) positioned precisely on their respective curves above. There are no perfect species or perfect stars, nor even necessarily average members within any category of complex system; nor are there likely to be exceptionless regularities or evolutionary "laws" in the real world. Explicit, singular values of $\Phi_m$ for individual complex systems are unlikely to pertain, rather only optimal ranges of $\Phi_m$ for each type of system. Nor should overlapping $\Phi_m$ values among nearly comparable complex systems cause concern; minor overlaps are common all along the master curve in Figure 2, much as might be expected, for example, when the simplest (dwarf) stars compare with the most complex (active) galaxies, or when advanced photosynthesizing sugarcane overlaps with some cold-blooded reptiles. Whether for stars and galaxies, or plants and animals, or society and technology, rare outliers, exceptions and overlaps are occasionally evident—indeed expected—among complex systems in an imperfect Universe.

*5.9. Cultural Complexity.* Some big historians (notably [3]) have expressed skepticism about pursuing cosmic evolution into the realm of worldly culture, claiming that the nature of complexity for human society and its built machines differs fundamentally from that of other systems in the Universe. They draw a subjective distinction between naturally evolving complexity and human-made "artificial" complexity, arguing that the former appears spontaneously (but it does not) whereas the latter is constructed by us and thus different (yet artificiality, like intentionality or directionality, are irrelevant in evolution). By contrast, I have always maintained that we, too, are a part of Nature, not apart from it; schemes that regard humankind outside of Nature, or worse atop Nature, are misguided. If we are to articulate a unified worldview for all complex systems observed throughout Nature, then we must objectively and consistently model each of them identically. To restate once more for clarifying emphasis, complex systems likely differ fundamentally not in kind, but only in degree—*i.e.*, degree of complexity manifesting ontological continuity. The critics' main anxiety is that cultural values of $\Phi_m$ often exceed those of humankind (*cf.*, Figures 8 and 9), and they are apparently unable or unwilling to accept that some culturally invented systems can be more complex than our own biological selves. However, technological devices were not built by Nature without intelligent beings, so it seems reasonable that some cultural systems' $\Phi_m$ values actually do sometimes exceed those of biological systems, just as life-forms outrank simpler physical systems; perhaps chance (and necessity) does favor the prepared mind [229]. Cultural evolution is a product of biological evolution, the former building upon the achievements of the latter. Provenance counts; networks of bodies and brains within the human web can build elaborate systems. And it is the rapid pace of cultural evolution, in addition to its ability to harness energy intensely, that makes cultural systems so remarkable. Accordingly, I expect cultural products to be typically more complex, and naturally so, than the biological systems that produce them; yet, within their range of variations (*cf.*, Sec. 5.8), not all necessarily are, such as pencils, mousetraps, and can openers that are relatively simple devices and not sustained systems per se, in fact once made begin decaying (*cf.*, Sec. 5.5). I am also comfortable with the empirical finding that some cultural systems, notably machines, computers, and cities that help in numerous ways to improve our health, wealth, and security are likely more complex than we are; jet aircraft operating in three-dimensions and computing extremely quickly may well be a hundred times more complex than a thinking mammalian being, as their $\Phi_m$ values imply. After all, it is the intricacies of our human brains and social networks that have made machines possible, so why should any machine—including vacuum cleaners and lawn mowers—be less complex or have, by design, smaller concentrated energy flows? Try gliding off a cliff with your body, cleaning a carpet with your brain, or even beating an iPhone at checkers; machines perform functions that biota cannot, often impressively so, and more rapidly too. Function also counts; flying high and computing fast are qualities that humans do not possess. This is not to say that cultural systems are smarter than we are; no claim links the complexity metric $\Phi_m$ to raw intelligence, rather only that some cultural systems are arguably more intricate and complicated, much as implied by the master graph in Figure 2. For big historians to declare that sentient, technological society is not analyzable in the same way as stars, galaxies, and life itself is tantamount to placing ourselves anthropocentrically in some special category or atop some exalted pedestal, raising the age-old spectre of mystical rulers and arrogant institutions. It would be as though Nature adheres to a universal concordance, creating all known systems in a single, unified, evolutionary way— but only until the big-history story reaches us, at which time society and our cultural inventions are alleged to be different, or artificial, or privileged. I reject such teleology, which has so often been detrimental to humankind during much of recorded history. My stance on cosmic evolution, in this review as well as in my decades-long research program, very much includes culture and civilization among all natural systems, indeed regards human society and our remarkable technology "on the same page" (as literally in Figure 2) alongside every type of complex system observed in the Universe. I urge caution when professing, egocentrically or for reasons of personal belief, that the complexity of social systems differs in kind from that of any other organized system. There is no objective evidence for humankind's speciality and no need to assert it subjectively.

*5.10. Sigmoidal curves.* Close examination of many of the graphs in this paper suggests that $\Phi_m$ often rises sharply for each type of complex system for only limited periods of time, after which the curves begin to turn over or flatten. Although caution is needed not to over-interpret these data, some (but not all) systems do slow their rate of complexification; they seem to follow a classic, sigmoidal, S-shaped curve—much as microbes do in a petri dish while replicating unsustainably or as human population is expected to plateau later this century. That is, $\Phi_m$ values for a whole array of physical, biological, and cultural systems first increase slowly and then more quickly during their individual evolutionary histories, eventually leveling off throughout the shaded area of Figure 2; if true, then the master curve of Figure 2 is probably the compound sum of multiple S-curves [230]. Note that $\Phi_m$ for viable, complex systems evidence no absolute decrease, rather merely lessened growth rates and S-shaped inflections as those systems apparently matured [231]. Some colleagues then conclude that $\Phi_m$ itself decreases, but it typically does not, at least not for surviving systems able to command optimal energy; others interpret it to mean complexity declines—yet that is also wrong. The rate of change of $\Phi_m$—which is itself a rate quantity—might eventually decrease, but that implies only that complexity's growth rate diminishes (calculus' first derivative), not necessarily the magnitude of complexity per se (second derivative). Ultimately most systems, including unstable stars, stressed species, and inept civilizations, do collapse when they can no longer sustain themselves by optimally managing their energy flows; such adverse fates, which are natural, common outcomes of cosmic evolution, are partly the subject of another study that explores practical applications of cosmic evolution to human society [22].

## 6. Summary

Physical, biological, and cultural evolution has produced a wide spectrum of complexity in Nature, each comprising an integral part of an all-inclusive, cosmic-evolutionary scenario of who we are and whence we came. Galaxies, stars, and planets, as well as life, society, and machines, play roles in a comprehensive story of ourselves, our world, and our Universe. For all these systems and many more, their dynamical steady states act as sources of novelty and innovation, taking advantage of random chance and lawful determinism to advance along the arrow of time toward greater complexity. Among myriad manifestations of order and organization on Earth and beyond, complex systems seem governed by common processes and properties, as though simple, underlying Platonic Forms pervade the cosmos.

Cosmic evolution is an extensive scientific narrative of changes, events, and processes that provide a quantitative basis for the study of big history during the past ~14 billion years, from big bang to humankind. It addresses the integrated topics of evolving systems and rising complexity, revealing how all known complex systems, from fusing stars and twirling galaxies to buzzing bees and redwood trees, are fundamentally related. And despite the patent messiness of much that surrounds us in Nature, it contends that evolution, when broadly conceived, potentially provides a unifying theme for much of modern science.

No purpose or plan is evident in the observed rise of universal complexity for those systems able to utilize optimally energy flowing through them; there is no evidence whatsoever that cosmic evolution obeys some grand design or intelligent designer. Nor is there any obvious progress either; we who study Nature incrementally progress in understanding while ambitiously deciphering this grand scientifically-based story, but no compelling evidence exists that evolution itself is progressive or directed (as in "movement toward a goal or destination"); cosmic evolution is an aimless, meandering process, partly facilitated by energy flowing through open, non-equilibrium, complex systems.

As a confirmed empirical materialist, my vocation is to critically observe Nature and to experimentally test theories about it—a mainstream application of the traditional scientific method, albeit in this case on behalf of a voluminous, interdisciplinary subject. Not that subjectivity is absent in science while practiced; rather, objectivity is eventually revealed only after much quantitative probing of qualitative ideas. Those ideas that pass the ultimate test of time endure—and those that do not are discarded; scientific hypotheses are subject to change, selection, and accumulation much like the many complex systems featured in this review article.

More than perhaps any other single operational factor, energy flow is a central leitmotif embedding all aspects of physical, biological, and cultural evolution. Energy flows in an expanding cosmos seem (at least partly) to dictate the emergence, maturity, and destiny of organized structures observed in Nature. In particular, energy rate density, $\Phi_m$, robustly contends as an unambiguous, quantitative measure of complexity, enabling detailed assessment of myriad ordered systems in like manner—a consistent empirical metric that gauges how over the vast course of natural history *in toto* some systems optimally commanded energy and survived, while others could not and perished. At all times and places in the Universe, physical laws apparently comprise an ultimate arbiter for Nature's many varied, complex systems, thereby guiding the origin and evolution of all material things.

Human society and its invented machines are among the most energy-rich systems known, hence plausibly the most complex systems yet encountered in the Universe. Cultural creations, bolstered by increased energy allocation as numerically tracked by rising $\Phi_m$ values, enable 21$^{st}$-century *H. sapiens* not only to adapt rapidly to our environment on Earth but also to manipulate it if desired, indeed to escape it if needed. Technological civilization and its prodigious energy usage arguably act as catalysts, speeding the course of cultural change, which like overarching cosmic evolution itself is unsettled and unpredictable. Yet societal complexification, which has decidedly bettered the quality of human life as measured by health, wealth, and security, inevitably grew—and continues to grow—at the expense of degraded environments and constant demands for yet more energy, which now powers humankind toward a fate unknown.

Earth is now in the balance. Our planet harbors a precarious collection of animate and inanimate localized systems amidst an intricate web of global energy flows. All these complex systems—whether non-humanly natural or humanly built—need to heed the laws of thermodynamics as unavoidable ground rules governing their existence. Consciousness, too, including societal intentions and technological decisions likely to dominate our actions for as long as our species endures, will likely require a broad evolutionary outlook, for only with awareness and appreciation of the bigger picture can we perhaps survive long enough to continue playing a role in our own cosmic-evolutionary worldview. All things considered, humanity, together with its society and its machines, might be among the minority of winning complex systems in Nature, continuing to make big history while advancing cautiously along the arrow of time.

## Acknowledgements

I thank numerous faculty and students at Harvard University and colleagues at Harvard-Smithsonian Center for Astrophysics for insightful discussions of the interdisciplinary topic of cosmic evolution. Challenging email exchanges with Fred Spier and David Baker have been especially helpful to me regarding big history scholarship, as were the detailed comments of an anonymous referee. This research has been supported over the past many years by the Sloan Foundation, Smithsonian Institution, National Aeronautics and Space Administration, National Science Foundation, and la Fondation Wright de Geneve.

## Conflicts of Interest

The author declares no conflicts of interest regarding publication of this paper.